\documentclass[psamsfonts]{amsart}

\numberwithin{equation}{section}

\usepackage{graphics}
\usepackage[dvipdf]{graphicx}
\usepackage{epstopdf}
\usepackage{epsfig}
\usepackage{url}

\newcommand \be{\begin{equation}}
\newcommand \bea{\begin{eqnarray}}
\newcommand \ee{\end{equation}}
\newcommand \eea{\end{eqnarray}}

\usepackage{epsf}

\usepackage{subfigure}
\usepackage{color}

\begin{document}

\title[Interplay between frictional sliding and damage
cracking]{Gravity-driven instabilities: interplay between state-and-velocity
  dependent\\ frictional sliding and stress corrosion damage cracking}

\markleft{J. FAILLETTAZ ET AL.}

\author{J. Faillettaz}
\address{VAW, ETH Zurich, Laboratory of Hydraulics, Hydrology and Glaciology\\Switzerland} 
\email{faillettaz@vaw.baug.ethz.ch}
\url{http://www.glaciology.ethz.ch}
\author{ D. Sornette}
\address{Department of Management, Technology and Economics, ETH Z\"urich}
\address{Department of Earth Sciences, ETH Z\"urich}
\address{Institute of Geophysics and Planetary Physics, UCLA}
\curraddr{Department of Management, Technology and Economics,\\ ETH Z\"urich\\Switzerland}
\email{ dsornette@ethz.ch}

\author{M. Funk}
\address{VAW, ETH Zurich, Laboratory of Hydraulics, Hydrology and Glaciology\\Switzerland}
\email{funk@vaw.baug.ethz.ch}

\thanks{
The authors acknowledge helpful discussions and exchanges with G. Ouillon.
}



\date{}

\begin{abstract}
We model the progressive maturation
of a heterogeneous mass towards a gravity-driven instability, characterized
by the competition between frictional sliding
and tension cracking, using array of slider blocks on an inclined basal surface, which interact
via elastic-brittle springs. A realistic state- and rate-dependent friction
law describes the block-surface interaction.
The inner material damage occurs via
stress corrosion. Three regimes, controlling the mass instability
and its precursory behavior, are classified as a function
of the ratio $T_c/T_f$ of two characteristic time
scales associated with internal damage/creep and with frictional sliding.
For $T_c/T_f \gg 1$, the whole
mass undergoes a series of internal stick and slip events, associated with an initial
slow average downward motion of the whole mass, and
progressively accelerates until a global coherent runaway is observed.
For $T_c/T_f \ll 1$, creep/damage occurs
sufficiently fast compared with nucleation of sliding, causing bonds to break, and the bottom part of the mass
undergoes a fragmentation process with the creation of a heterogeneous population of
sliding blocks. For the intermediate regime $T_c/T_f  \sim 1$, a macroscopic crack nucleates
and propagates along the location of the largest curvature
associated with the change of slope from the stable frictional state
in the upper part to the unstable frictional sliding state in the lower part.
The other important parameter is the Young modulus $Y$
which controls the correlation length of displacements in the system.
\end{abstract}

\maketitle

\section{Introduction}
Gravity-driven instabilities include landslides, mountain collapses, rockfalls,
ice mass break-off and snow avalanches. They pose a considerable risk to
mountain communities, real-estate development, tourist activities and hydropower
energy generation. Gravity-driven instabilities are the most widespread natural
hazard on Earth. Most of these consist of slumping masses of soil, too small to be lethal but
costly in term of  property damage, especially when they are close to roads, railway lines and
built-up areas. At the other extreme is the infrequent occurrence of the
collapse of a mountainside which can release the energy equivalent of a major volcanic
eruption or earthquake within seconds. Such collapses are among the most
powerful hazards in nature and can wreak devastation over tens of square
kilometers. In the US and Europe in particular, gravity-driven instabilities at
all scales are particularly significant and cause billions of dollars and euros in
damage each year.

Gravity-driven instabilities occur in a wide variety of geomechanical contexts,
geological and structural settings, and as a response to various loading and
triggering processes. They are often associated with other major natural
disasters such as earthquakes, floods and volcanic eruptions. Some slope
instabilities involve the rupture and collapse of entire mountains, with
devastating consequences. One of the most spectacular rockslides that ever
occurred in the world is the K\"ofels rockslide [Erismann and Abele, 2001], in
which more than $3\;\rm km^3$ of rock was released in a short time span. This event
occurred approximately 9,600 years ago. Many other large and giant
historic and prehistoric rockslides have been described by Heim [1932], Abele
[1974] and Erismann and Abele [2001]. Recent spectacular landslides in the
European Alps include the Vaiont disaster, which resulted in about 2,000 casualties
when a $0.3\; \rm km^3$ sliding mass fell into a hydro-electric power reservoir in
1963 [M\"uller, 1964]. Another example is the Val Pola slide, damming up the valley in 1987 [Azzoni
et al., 1992]. Much can also be learned from frequent and historically reported
giant and rapid landslides on the slopes of many volcanoes [Moore et al., 1994;
McMurtry et al., 1999; Clouard et al., 2001]. These are often associated
with either caldera collapse [Hurlimann et al., 2000] or volcanic eruption
[Voight, 1988; Heinrich et al, 1999]. Earthquakes are other triggering
processes, which have to be considered.

The present paper develops a model of the progressive maturation
of a mass towards a gravity-driven instability, which combines
basal sliding and cracking. Our hypothesis is that
gravity-driven ruptures in natural heterogeneous material are characterized by a common
triggering mechanism resulting from a competition between frictional sliding
and tension cracking. Heterogeneity of material properties and dynamical
interaction seem to have a significant influence on the global behavior.
Our main goal is thus to understand
the role of the competition or interplay between the two physical 
processes of sliding and tension cracking in the early stages of an instability. 
The run-up to the
sliding instabilities can be described by applying a modern constitutive law of 
state-and-velocity dependent friction. This means that solid friction is not used as a
parameter but as a process evolving with the concentration of deformation and
properties of sliding interfaces. Cracking and fragmentation in the mass is accounted
for by using realistic laws for the progressive damage accumulation
via stress corrosion and other thermally activated processes aided by stress.
In one sense, the present paper can be considered as an important
extension of [Helmstetter et al., 2004; Sornette et al., 2004], who
modeled the potentially unstable mass by means of a single rigid slider block
interacting with a inclined basal surface via solid friction. 
These authors showed that the use of a realistic 
state- and rate-dependent friction law established in the laboratory 
could provide a reasonable starting point for describing the
empirical displacement and velocity data preceding landslides.
Here, we use an array of such slider blocks on an inclined
(and curved) basal surface, which interact 
via elastic-brittle springs. The springs, which model the inner 
material properties of the mass, also undergo damage, eventually
leading to failure.  By combining both sliding and damage processes
in a single coherent model, this paper contributes to the literature by providing
\begin{itemize}
\item a better understanding of the relative role of and interplay between the 
frictional and rupture processes, 
\item a quantification of the transition from
slow stable landsliding to unstable and fast catastrophes events, 
\item a theory of potential precursors to catastrophic slope failure, 
\item a model to exploit the
increasing availability of continuous monitoring by geodetic, GPS, SAR
measurements as well as geophysical measurements to obtain advanced warnings,
\item a framework to assess
the impact of varying climatic conditions and/or external forcing (rain, snow,
passing seismic waves, etc.),  and
\item a framework for developing strategies to mitigate the hazard resulting from
gravity-driven instabilities. 
\end{itemize}

In Section 2, we provide a summary of previous works related to gravity-driven instabilities,
and focus particularly on the different strategies employed to investigate stability. Section 3 is
devoted to a detailed description of the proposed model. Section 4 presents an
application of this numerical model
to a hanging glacier and a comparison between numerical and experimental
results. Conclusions are presented in Section 5.

\section{Summary of previous related works}

At the longest time scales, gravity-driven instabilities have been
studied historically by the method of slope stability analysis 
developed in civil engineering, which was developed to ensure the safety of man-made structures, such
as dams and bridges. This method usually divides a profile view of the mass of
interest into a series of slices, and calculates the average factor of safety for
each of these slices, using a limit equilibrium method [e.g. Hoek and Bray, 1981;
Blake et al., 2002]. The factor of safety is defined as the ratio of the
maximum retaining force to the driving forces. Geomechanical data and properties
are inserted in finite elements or discrete elements of numerical codes, which are
run to predict the possible departure from static equilibrium or the distance to a failure
threshold. This is the basis of landslide hazard maps, using safety factors
greater than $1$. In this modeling strategy, the focus is on the recognition of
time-independent landslide-prone areas. Such an approach is similar to the
practice in seismology called ``time-independent hazard,'' where
earthquake-prone areas are located relative to active faults, for instance, while
the prediction of a single, individual earthquake is recognized to be much more
difficult, if not unattainable. This ``time-independent hazard'' essentially
amounts to assuming that gravity-driven instabilities result from a random
(Poisson) process in time, and uses geomechanical modeling to constrain the
future long-term hazard. In this model, failure appears to occur without
warning, probably because earlier movements have passed unnoticed. The
approaches in terms of a factor of safety do not address the possible
preparatory stages leading to the catastrophic collapse.

In contrast, ``time-dependent hazard'' would accept a degree of predictability
in the process, because the hazard associated
with the potential triggering of gravity-driven instabilities varies with
time, perhaps in association with varying external factors (rain, snow,
volcanic). For instance, accelerated motions have been linked to pore pressure
changes [e.g. Vangenuchten and Derijke, 1989; Van Asch et al., 1999], leading to
an instability when the gravitational pull on a slope exceeds the resistance at
a particular subsurface level. Since pore pressure acts at the level of
submicroscopic to macroscopic discontinuities, which themselves control the
global friction coefficient, circulating water can hasten chemical alteration of
the interface roughness, and pore pressure itself can force adjacent surfaces
apart. Both effects can produce a reduction in the friction coefficient which
leads, when constant loading applies, to accelerating movements [Kuhn and
Mitchell, 1993]. However, this explanation has not lead to a quantitative method
for forecasting slope movement.

The next level in the hierarchy of models would be ``gravity-driven instability
forecasting,'' which would require significantly better understanding to enable the prediction of some of the features of an impending 
catastrophe, usually
on the basis of the observation of precursory signals. Practical difficulties
include identifying and measuring reliable, unambiguous precursors, and the
acceptance of an inherent proportion of missed events or false alarms. Let us
also mention the approach developed by [Iverson, 1985; 1986a; 1986b; 1993; Iverson and
Reid, 1992] based on coupled pressure-dependent plastic yield and nonlinear
viscous flow deformations (see also [Iverson and Reid, 1992]). According to
the theory of Iverson et
al., the plastic yield is a generalization of the Coulomb criterion to
three-dimensional stress states in the spirit of Rice [1975], and the effect of
pore-water pressures is accounted for by the use of the Terzaghi ``effective''
stresses. This theory embodies as special cases models of creeping, slumping,
sliding and flowing types of slope deformation. The underlying equations remain
fundamentally based on continuum mechanics, which differentiates it from our
approach which focuses on the dominating role of sliding interfaces/bands and their
rheology controlled by discrete asperities, joints and blocks.

There exists
a common phenomenology prior to rupture, which is observed in many different
natural materials and geophysical set-ups. In particular, 
there are well-documented cases of precursory signals, showing accelerating slip-over time scales of weeks to decades. 
We mentioned before the Vaiont landslide on the Mt Toc slope in the
Dolomite region in the Italian Alps, which
was the catastrophic culmination of an accelerated slope
velocity (see for instance [Helmstetter et al., 2004; Sornette et al., 2004]
for the data and a recent analysis in terms of accelerated precursory sliding velocity).
More precisely, gravity-driven instabilities often exhibit a finite time singularity
characterized by a power law behavior of some variable
(such as displacement or deformation, velocity of the instable mass or acoustic
emissions) as a function of time.
This behavior is quite common in a wide class of nonlinear
process, including gravity-driven instability such as rockfall [Amitrano et
al., 2005], landslides [Helmstetter et al., 2004; Sornette et al., 2004],
break-off of ice chunks from hanging glaciers [Flotron, 1977; R\"othlisberger,
1981; L\"uthi, 2003; Pralong and Funk, 2005, Pralong et al., 2005], 
but also for  earthquakes [Bufe and Varnes, 1993; 
Bowman et al., 1998; Jaum\'e and Sykes, 1999; Sammis and Sornette, 2002], and
volcanic eruption [Voight, 1988]. Analogous behaviors
have also been associated with financial crashes,  with 
the dynamics of the economy and of 
population [Johansen and Sornette, 2001; Ide and Sornette 2002].
In other cases which exhibited accelerated deformations , such as La Clapi\`ere mountain [Helmstetter et al., 2004]
in the southern Alps near St-Etienne-de-Tin\'ee, France, the sliding can undergo a transition
towards a slower more stable regime. While only a few such cases
have been monitored in the past, modern monitoring techniques are bound to
provide a wealth of new quantitative observations based on GPS and SAR
(synthetic aperture radar) technology to map the surface velocity field
[Mantovani et al., 1996; Parise, 2001; Fruneau et al., 1996; Malet et al., 2002;
Berardino et al., 2003; Coe et al., 2003; Colesanti et al., 2003; Mora et al.,
2003; Wasowski and Singhroy, 2003] and seismic monitoring of slide quake
activity [Gomberg et al., 1995; Xu et al., 1996; Rousseau, 1999; Caplan-Auerbach
et al., 2001].

Other studies proposed that rates of gravity-driven mass movements are controlled
by microscopic slow cracking and, when a major failure plane is
developed, the abrupt decrease in shear resistance may provide a 
sufficiently large force imbalance to trigger a catastrophic slope rupture
[Kilburn and Petley, 2003]. Such a mechanism, with a proper law of input of 
new cracks, may reproduce the acceleration
preceding of the collapse that occurred at
Vaiont, Mt Toc, Italy [Kilburn and Petley, 2003].

An alternative modeling strategy consists in viewing the
accelerating displacement of the slope prior to the collapse as the
final stage of the tertiary creep preceding
failure [Saito and Uezawa, 1961; Saito, 1965; 1969; 
Voight and Kennedy, 1979; Voight, 1988; 1989; 1990; Fukuzono, 2000;
Kuhn and Mitchell, 1993]. 
Further progress in exploring the relevance of this mechanism
requires a reasonable knowledge of the geology of the sliding surfaces,
their stress-strain history, the mode of failure and the time-dependent shear
strength along the surface of failure
[Bhandari, 1988]. Unfortunately, this range of information is not usually available. 
This mechanism is therefore used mainly as a justification for
the establishment of empirical criteria of impending landslide
instability. 

Observations of landslides have been quantified by
what amounts to a scaling law relating
the slip acceleration $d\dot{\delta}/dt$ 
to the slip velocity $\dot{\delta}$ according to
the following equation
[Voight, 1990; Voight and Kennedy, 1979; Petley et al., 2002] 
\be
d\dot{\delta}/dt = A \dot{\delta}^{\alpha}~,
\label{mgnlsa}
\ee
where $A$ and $\alpha$ are
empirical constants. For $\alpha>1$, this relationship predicts
a divergence of the sliding velocity in finite time at some critical time $t_f$,
which is determined by the parameters $A$ and $\alpha$ and the
initial value of the velocity.
The mathematical divergence is of course not to be taken literally: it signals 
a bifurcation from accelerated creep to complete slope instability for which
inertia is no longer negligible and other processes such as liquefaction
(not necessarily involving fluids) come into play. 
Several cases have been quantified ex-post with this law,
usually using $\alpha=2$, by plotting the
time $t_f-t$ to failure as a function of the inverse of the creep velocity
[Bhandari, 1988]. In fact, the solution of (\ref{mgnlsa}) for $\alpha > 1$ 
can be expressed as
\be
t_f-t \sim 1/ \dot{\delta}^{1 \over \alpha-1}~.
\label{jmgjkla}
\ee
These fits suggest that it might be possible to forecast
impending landslides by recording accelerated precursory slope
displacements. As a matter of fact, for Mont Toc, Italy, Vajont landslide, 
Voight [1988] mentioned that
a prediction of the failure date could have been made 
more than 10 days before the actual failure by using (\ref{jmgjkla}) with 
$\alpha=2$. The physically based friction model [Helmstetter et al., 2004; Sornette et al., 2004] 
which avoids such a priori assumption has confirmed this claim.

\begin{figure*}[h]
\centerline{
\includegraphics[width=39pc]{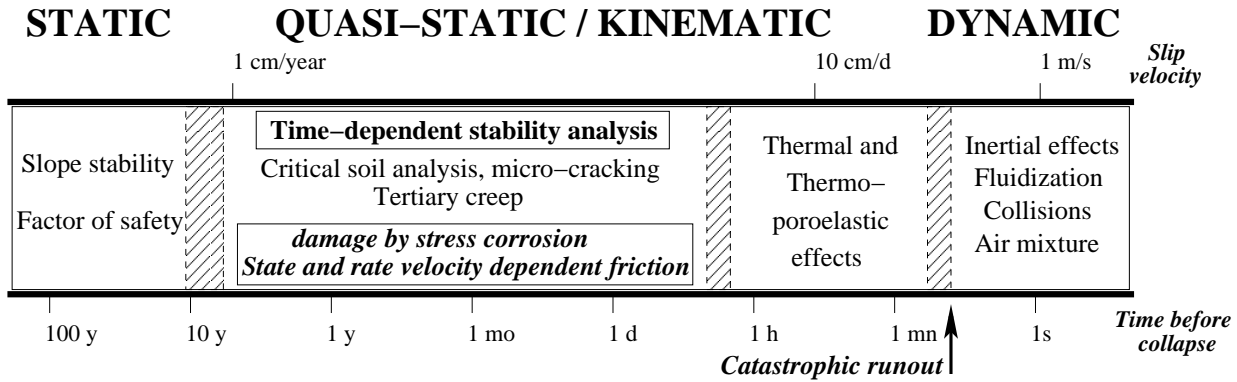}}
\caption{Simplified synopsis of the range of time scales, 
the associated sliding velocities and the major processes involved in the
evolution of slopes (see for example [Brunsden, 1999] for a review). Our model
is applied to the intermediate range from years to tens of minutes before
the catastrophic collapse, as indicated
in boldface. This graph
was prepared with the help of G. Ouillon.
}
\label{timescales}
\end{figure*}

Voight [1988; 1989] proposed that relation~(\ref{mgnlsa}),
which generalizes damage mechanics laws [Rabotnov, 1980; Gluzman
and Sornette, 2001], can be used with other variables (including
strain and/or seismic energy release) for a large variety of materials
and loading conditions [Sammis and Sornette, 2002]. 
Equation~(\ref{mgnlsa})
seems to apply as well to diverse types of landslides occurring in
rock, soil and ice, including first-time and reactivated slides [Voight, 1988].
It may be seen as a special case of a
general expression for failure [Voight, 1988; 1989; Voight and Cornelius,
1991]. Recently, such time-to-failure laws have been interpreted
as resulting from cooperative critical phenomena and
have been applied to the prediction of failure of 
heterogeneous composite materials [Anifrani et al., 1995]
and to precursory increase of seismic activity prior to main shocks
[Sornette and Sammis, 1995; Jaume and Sykes, 1999; Sammis and Sornette, 2002].
See also [Sornette, 2002] for extensions to other fields.

The law~(\ref{mgnlsa})
may apply over time scales from years to tens of minutes before the run-out, as
summarized in Figure~\ref{timescales}. Our model provides a 
general conceptual and operational framework to rationalize the law~(\ref{mgnlsa}).
One of the strengths of our approach
is to be able to tackle in the same conceptual framework
both stable slope evolutions as well as unstable regimes culminating in a runout.
Our approach applies to any landslides, irrespective of their final
catastrophic evolution in the runout regime, because
we focus on the quasi-static regime in which the slope can still be
analyzed as a system of interacting sliding blocks. While we model
the progressive damage and rupture between sliding blocks, we
do not describe the final stage of the dynamic runout during which
most large landslide disintegrate in rapid motion in a collection of blocks as they move
down the slope (as an example of this, see [Campbell 1989; 1990;
Cleary and Campbell, 1993; Campbell et al., 1995]). This final dynamic regime
includes additional physical mechanisms that we do not account for:
(i) heat-generated pore pressure inside a rapidly
deforming shear band [Voight and Faust, 1982a,b], (ii)
thermo-poro-mechanical softening of the soil [Vardoulakis, 2002a,b]
and (ii) shear-induced pore dilation which leads to unstability
when sediments are dilated to their critical-state porosity 
(see for instance [Moore and Iverson, 2002]). These models predict
exponential accelerations only in the last tens of seconds when the 
sliding velocity is sufficiently large. Finally, the runout itself
is characterized by further mechanisms associated with velocities
reaching such large values (meter/s to tens of meters/s)
that inertial effects are no longer negligible (see Figure~\ref{timescales}). 
These two times scales are not addressed in our approach, which 
focuses on the stability and possible destabilizations of
slopes over time scales from years to tens of minutes.

\section{Geometry and physical inputs of the model \label{bvhbcew}}

\subsection{Main elements of the model}

To tackle the challenge of modeling
the interplay between the two physical 
processes of sliding and tension cracking in the preparatory stages of a
gravity-driven instability, we proposed a discrete set-up made
of block masses linked to each other by elastic springs.
The blocks slid on a basal inclined surface and interacted via direct elastic coupling
to the neighboring blocks.

A toy model embodying the principles of competition between
subsurface slip and the possible growth of cracks decoupling adjacent blocks was
developed by Andersen et al. [1997]
and Leung and Andersen [1997].
Many other investigators have studied spring-block models, but most were derived
from the Burridge-Knopoff prototype having elastic coupling with an upper rigid plate,
as a simple representation of the tectonic driving which lead to earthquakes.
Note that such elastic coupling to an upper plane 
entails the wrong large scale elastic limit for gravity-driven masses. In contrast, the blocks in
the Andersen et al. [1997]'s model are only connected to nearest
neighbors and are in frictional contact with a substrate. Meakin and
collaborators [Meakin, 1991] have also studied a similar model in the context of
drying patterns.

Using blocks as representative discrete interacting elements 
provides a reasonable starting point for describing the processes occurring
before the complete destabilization of the slope. Actually, 
a model with just one block was used by  [Heim, 1932; Korner, 1976; Eisbacher,
1979; Davis et al., 1990]. 
In these works, the friction coefficient between the rigid block and 
surface is usually imposed as a constant, or just either slip- or
velocity-dependent (but not slip- and velocity-dependent). 
A constant solid friction coefficient (Mohr-Coulomb law) is often taken
to simulate mass over bedrock sliding. A slip-dependent
friction coefficient model is taken to simulate the
yield-plastic behavior of a brittle material beyond 
the maximum of its strain-stress characteristics.
For rock avalanches, Eisbacher [1979] suggested that
the evolution from a static to a dynamic friction coefficient 
is induced by the emergence of a basal gauge. 
Studies using a velocity-dependent friction coefficient have focused mostly on the establishment of empirical relationships
between shear stress $\tau$ and block velocity $v$, 
such as $v \sim \exp [a \tau]$ [Davis et al., 1990] or
$v \sim \tau^{1/2}$ [Korner, 1976], however without a 
definite understanding of the possible mechanism. 
Muller and Martel [2000] also point out that near the surface, principal
stresses are either normal or parallel to the local slope; thus
shear stress on planes parallel to the slope must be small.
This implies that slip must initiate on pre-existing planes of weakness.

Our present model improves on these models
and on the multi-block model of Andersen et al. [1997]
and Leung and Andersen [1997] in two ways. First, we use a
state-and-velocity weakening friction law instead of 
a constant (or just state- or velocity-weakening) solid friction coefficient. Second, rather than a static
threshold for the spring failures, we model the 
progressive damage accumulation
via stress corrosion and other thermally activated processes aided by stress.
Both improvements make the numerical simulations significantly
more involved but present the advantage of embodying rather well
the known empirical phenomenology of sliding and damage processes.
Adding the state and velocity-dependent friction law and time-dependent
damage processes allows us to model rather faithfully the 
interplay between sliding, cracking between blocks and the overall
self-organizing of the system of blocks. This will allow us to 
investigate the following questions:
what are the conditions under which cracking (disconnection
between blocks) can stabilize sliding? What is the effect of heterogeneity along
the slope? Can we construct a simple set of blocks and their interactions that
can reproduce the complex history of some slopes, such as the
case studies presented in Section~\ref{thjetrobj3r}.

\begin{figure}
\centerline{
\includegraphics[width=20pc]{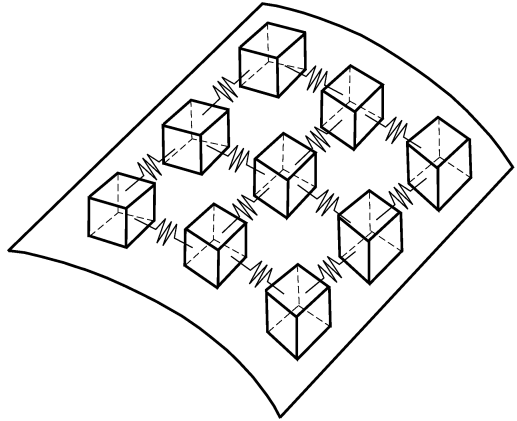}}
\caption{Illustration of the model constisting of spring-blocks resting on an inclined slope.
The blocks lie on an inclined curved surface and gravity is the driving force. Only
a small subset of the spring-block system is shown here.
}
\label{blocks}
\end{figure}

The geometry of the system of blocks interacting via springs and with 
a basal surface is depicted in Figure~\ref{blocks}. 
The model includes the following characteristics:
\begin{enumerate}
\item frictional sliding on the ground or between layers,
\item heterogeneity of basal properties,
\item possible tension rupture by accumulation of damage,
\item dynamical interactions of damage or cracks along the sliding layer,
\item geometry and boundary conditions,
\item interplay between frictional sliding and cracking.
\end{enumerate}

We now turn to the specification of the two key ingredients, the friction and
damage laws, that are applied to blocks and bonds respectively.

\subsection{Friction law between the discrete blocks and the basal surface
\label{frictionaa}}

\subsubsection{State- and rate-dependent solid friction}

State- and velocity-dependent
friction laws have been established on the basis of
numerous laboratory experiments (see,
for instance, [Scholz, 2002; 1998; Marone, 1998; Gomberg et al., 2000] for reviews). 
Friction laws have been investigated in the laboratory using mainly 
strain-controlled tests in which, for instance, motion of a slider block
is driven by controlling its velocity. This strain-controlled set-up
is thought to be relevant to faults which are generally assumed to be
driven by far-field forces transmitted via elastic and inelastic deformations. 
In contrast, gravity-driven masses move chiefly in response to changes in the mix of
internal body forces imposed by gravity and pore-pressure fields. In this context,
stress-controlled experiments are more relevant. Dieterich [1994] has shown that
his strain-controlled friction law does apply to stress-controlled
interactions between ruptures. Many other works have used these friction law
in stress-controlled situations
(see e.g. [Lapusta et al., 2000; Ben-Zion and Rice, 1997]). 
Experiments have also confirmed that the state- and velocity-dependent friction
laws established with strain-controlled experiment apply equally well
(with a slight modification with regularization) to
stress-controlled situations [Prakash and Clifton, 1993; Prakash, 1998;
Ben-Zion, 2001; Berger, 2002]. 

Analogies between landslide faults and tectonic faults
[Gomberg et al., 1995] and the relevance of a solid friction criterion 
for pre-existing slopes with weak layers or geologic discontinuities, with
bedding planes, landslide slip-surfaces, faults, or joints, or for existing landslides, suggest
that the improved description of the friction coefficient that we
use here be viewed as an a element essential to the understanding of slope evolutions.
To our knowledge, Helmstetter et al. [2004] and Sornette et al. [2004] were the
first to explore quantitatively the analogy between sliding
rupture and earthquakes and to use the physics of state- and velocity-dependent
friction to model gravity-driven instabilities and their precursory phases
(see [Brabb, 1991;  Cruden, 1997; Guzetti et al., 1999; Howell et al., 1999;
Al-Homoud and Tahtamoni, 2001] for a compilation of cases). 
Earlier, Chau [1995, 1999] also used the state- and rate-dependent
friction law~(\ref{vxcxxzx}) given below to model gravity-driven
instabilities. He transformed the 
problem into a formal nonlinear stability analysis of the type found in
mathematical theory of dynamical systems, but no physical application
was described.  In particular, Chau's analysis missed
the existence of the finite-time singular behavior found also empirically,
as discussed above and recalled below.

The key concept underlying the state- and rate-dependent friction law
is that the friction coefficient $\mu$ between a block and the inclined basal
surface supporting it evolves continuously
as a function of cumulative slip $\delta$, slip velocity $\dot{\delta}$ and time, as shown by
experiments carried over a large range of time scales, under static or sliding
conditions over a large range of velocities, for a wealth of different materials.
The version of the rate/state-variable constitutive law, currently most
accepted as being in reasonable agreement with experimental 
data on solid friction, is known as the Dieterich-Ruina law 
[Dietrich, 1994]:
\be
\mu(\dot{\delta}, \theta) = \mu_0 + A \ln {\dot{\delta} \over \dot{\delta}_0} + B \ln 
{\theta \over \theta_0}~,
\label{vxcxxzx}
\ee
where the state variable $\theta$ is usually interpreted as the
proportional to the surface
of contact between asperities of the two surfaces. The constant $\mu_0$ is the
friction coefficient for a sliding velocity $\dot{\delta}_0$
and a state variable $\theta_0$. $A$ and $B$ are coefficients depending on material properties.
Expression~(\ref{vxcxxzx}) as written is not appropriate for very low
or vanishing values of sliding velocities and a simple regularization
near $\dot{\delta}=0$ can be performed, that is
motivated by Arrhenius thermal activation of
creep at asperity contacts [Ben-Zion, 2003].
The time evolution of the state variable $\theta$ is described by
\be
{{\rm d}\theta \over {\rm d}t} = 1 - {\theta \dot{\delta} \over 
D_c}~, 
\label{qwertyu}
\ee
where $D_c$ is a characteristic slip distance, usually interpreted as
the typical size of asperities. We note that (\ref{qwertyu})
can be rewritten
\be
{{\rm d}\theta \over {\rm d}\delta} = {1 \over \dot{\delta}}
- {\theta \over D_c}~. \label{qweadasgatyu}
\ee

For small or zero sliding velocity, $\theta$ grows linearly with
time. Using the above-mentioned regularization [Ben-Zion, 2003], this
gives a logarithmic strengthening of $\mu$.
For a steady-state non-zero velocity, we have [Scholz, 1998]
$\mu = {\hat \mu}_0 + (A-B) \ln [\dot{\delta} / \dot{\delta}_0]$,
where ${\hat \mu}_0 = \mu_0 + B \ln {D_c \over \theta_0 \dot{\delta}_0}$.
Thus, the derivative of the steady-state friction coefficient with
respect to the logarithm of the reduced slip velocity is $A-B$. If
$A>B$, this derivative is positive: friction increases with slip velocity
and the system is stable as more resistance occurs which tends to react against
the increasing velocity. In contrast, for $A<B$, friction exhibits the
phenomenon of velocity-weakening and is unstable.

The primary parameter that determines stability, $A-B$, is a
material property. For instance, for granite, $A-B$ is negative
at low temperatures (unstable) and becomes positive (stable) for temperatures above
about 300$^\circ$C. In general, for low-porosity
crystalline rocks, the transition from negative to positive $A-B$
corresponds to a change from elastic-brittle deformation to crystal
plasticity in the micro-mechanics of friction [Scholz, 1998].
For the application to gravity-driven instabilities, we should in addition consider that
sliding surfaces are not only contacts of bare rock surfaces: they
are usually lined with wear detritus, called cataclasite or gouge in the case of faults or joints.
The shearing of such granular material involves an additional
hardening mechanism (involving dilatancy), which tends to make
$A-B$ more positive. For such materials, $A-B$ is positive when
the material is poorly consolidated, but decreases at elevated
pressure and temperature as the material becomes lithified.
See also Section 2.4 of Scholz's book [Scholz, 2002] and the discussion below.
The friction law (\ref{vxcxxzx}) with (\ref{qwertyu}) accounts for
the fundamental properties of a broad range of surfaces in contact, 
namely that they strengthen (age)
logarithmically when aging at rest, and tend to weaken (rejuvenate) when sliding.

\subsubsection{Frictional dynamics}

Let us first consider a single block and the base in which it is encased. The block represents
the discrete element of the slope which may be potentially unstable.
The two main parameters
controlling the stability of the block are the angle $\phi$ between 
the surface on which the block stands and the horizontal and the solid friction 
coefficient $\mu$. The block exerts
stresses that are normal ($\sigma$)
as well as tangential ($\tau$) to this surface of contact.
The angle $\phi$ controls the ratio of the shear
over normal stress: $\tan \phi = \tau/\sigma$. 
In a first step, we assume for simplicity that the usual
solid friction law $\tau = \mu \sigma$ holds for all times.
Four possible regimes are found, that depend on the ratio $B/A$ of 
two parameters $A$ and $B$ of the rate and state friction law and on the initial
frictional state of the sliding surfaces characterized by a
reduced parameter 
\be
X_i=  (S\; \theta_0)^{1/(1-B/A)}~{\theta \over \theta_0}~, ~~~~~\;\;\;{\rm for}~A \neq B~, 
\ee
where
\be
S \equiv {\dot{\delta}_0  \over D_c}~ e^{{\tau \over \sigma}-\mu_0 \over A}
\label{mmjdl}
\ee
depends on the material properties but not on the initial conditions,
and $\theta_0$ represents the initial value of the state variable.

Helmstetter et al. [2004] have shown that 
two regimes among them can account for an acceleration of the displacement.
For  $B/A>1$ (velocity weakening) and $X_i<1$,
the slider block exhibits an unstable acceleration
leading to a finite-time singularity of the displacement
and of the velocity $v \sim 1/(t_f-t)$, thus rationalizing  {\it Voight}'s 
empirical law discussed in Section 2.
An acceleration of the displacement can also be reproduced in the
velocity  strengthening regime, for $B/A<1$ and $X_i>1$. In this case,
the acceleration of the displacement evolves toward a stable sliding with
a constant sliding velocity.
The two others cases ($B/A<1$ and $X_i<1$, and $B/A>1$ and $X_i>1$) give
a deceleration of the displacement.

The case $B=A$ is of special interest because it retrieves the main
qualitative features of the two classes, and also because, 
empirically, $A$ is very close to $B$. It may be natural to assume that 
$A=B$ to remove the need for one parameter and ensure more robust
results. In the sequel, we use this approximation  $A \approx B$.
In this case, expression~(\ref{qwertyu}) renders
\be
{{\rm d} \theta \over {\rm d} t} = 1-S\; \theta_0~.
\ee
If $S\;~ \theta_0>1$ and is a constant, $\theta$ decays
linearly and reaches $0$ in finite time. This retrieves the finite-time
singularity, with the slip velocity diverging as $1/(t_f-t)$ corresponding
to a logarithmic singularity of the cumulative slip.
If $S\;~ \theta_0<1$ and is a constant, $\theta$ increases linearly with time. As a consequence,
the slip velocity decays as $\dot{\delta} \sim 1/t$ at large times
and the cumulative slip grows asymptotically logarithmically
as $\ln t$. This corresponds to a standard plastic hardening behavior.

Appendix A calculates the critical time $t_f$ (for $S \theta_0>1$)
as a function of the total shear force
$T \equiv \Arrowvert \sum \vec F_{\rm bond}\;-\;T_{\rm weight} \vec x \Arrowvert$ 
and the normal force $N \equiv N_{\rm weight}$ exerted
on a given block, where $\vec F_{\rm bond}$ is the force exerted by a neighboring 
spring bond, $N_{\rm weight}$ and $T_{\rm weight}$ are the 
normal and tangential forces due to the weight of the block. These forces
enter into the value of $S$ defined by (\ref{mmjdl}) via $\tau$ and $\sigma$.
From the definition of a solid friction coefficient ($\mu = \frac{T}{N}$), the 
critical time $t_f$ signaling the transition from 
a subcritical sliding to the dynamical inertial sliding is, for $\mu>\mu_0$,
\be
t_f\;=\;\frac{\theta_0}{\exp(\frac{\mu-\mu_0}{A})-1}~,
\label{kthwhlhw}
\ee
while $t_f \to \infty$ for $\mu \leq \mu_0$.

\subsubsection{General algorithm of the frictional processes}

The simulation of the frictional process for each given block proceeds as follows:
\begin{enumerate}
\item A given configuration of blocks and spring tensions determines
the value of  $T \equiv \Arrowvert \sum \vec F_{\rm bond}\;-\;T_{\rm weight} \vec x \Arrowvert$ 
and $N \equiv N_{\rm weight}$ for each block,
and therefore of their solid friction coefficient $\mu$ equal to the ratio $T/N$.

\item  Knowing $\mu$ for a given block and with the other material parameters
$\theta_0, \mu_0$ and $A$ for that block, the time $t_f$  for the transition
to the dynamical sliding regime for that block is calculated with
expression~(\ref{kthwhlhw}). The formula~(\ref{kthwhlhw}) for 
$t_f$ gives, therefore, the waiting time until the next slide of that block. 

\item When the block undergoes a transition into the dynamical sliding regime at time $t_f$, 
its subsequent dynamics is taken to obey Newton's law. 

\item Following other investigators 
[Farkas et al., 2005; Persson, 1993], 
we assume that, when the dynamical sliding phase
starts, the friction coefficient decreases abruptly from its value attained at time $t_f$
to a smaller dynamical value $\mu_{\rm dyn}$ independent of
the sliding velocity.  Generally, the ratio between the kinetic and static friction coefficients
is bounded as $1/2  \le \frac{\mu_{dynamic}}{\mu_{static}} \le 1$.
The choice $\mu_{\rm dyn}\;=\;\frac{2}{3}\mu(t_f)$ corresponds
to the harmonic mean of the lower and upper bounds and is taken as 
a reasonable representative of the dispersed set of values reported empirically.
For the sake of simplicity, we impose this value for the 
dynamical friction coefficient in the dynamical regime.

\item The dynamical slide of the block goes on as long as the velocity
of the block remains positive. When its velocity first reaches zero, we assume
that the block stops. 
To account for the heterogeneity and roughness of the sliding surface,
we assume that the state variable $\theta_0$ is reset to a new random value after
the dynamical sliding stops. This random value
is taken to reflect the characteristics of the new asperities 
constituting the fresh surface of contact.
For the sake of simplicity, we use the same parameters $\dot{\delta_0}$ and $D_c$, which
we assume identical over all blocks and constant as a function of time. 

\item After a dynamical slide, the forces exerted by the springs that connected
the block to its neighbors are updated, as is the new gravitational
force (if the basal surface has a curvature), the new value of $\mu$ is 
obtained, the time counter for frictional creep is reset to zero, and a new process of slow frictional creep
develops over the new waiting time $t_f$, that is, in general, different from the 
previous one.

\end{enumerate}

\subsection{Creep and damage process leading to tensional rupture in bonds \label{hjjbgmfqq}}

We now turn to the description of the progressive damage
of the bonds between the blocks, which represents the possible formation
of cracks and the occurrence of fragmentation in the mass body.

\subsubsection{Stress corrosion and creep law  \label{tbhothjw}}

It is generally known that any material subjected to a stress, constant or not,
undergoes time-dependent deformation, known as creep. 
In creep, the stress is less than the mechanical strength of the material, but 
the material eventually reaches a critical state at some time $t_c$ at which a global 
catastrophic rupture occurs. By waiting a sufficiently long time, the cumulative damage building up
during the deformation under stress may finally end up in a catastrophic rupture. 
Creep is all the more important, the greater the 
applied stress and the higher the temperature. The time needed to reach
rupture under creep is controlled by the tensional versus the compressive
nature of the stress and its magnitude, by the temperature and the
microstructure of the material.  
Creep rupture phenomena
have long been investigated through direct experiments 
[Liu and Ross, 1996; Guarino et al., 2002; Lockner, 1998], and described on
the basis of different models 
[Miguel et al., 2002, Ciliberto et al., 2001; Kun et al., 2003; Hidalgo et al., 2002;
Main, 2000; Politi et al., 2002; Pradhan and  Chakrabarti, 2003; Turcotte et al., 2003;
Shcherbakov and Turcotte, 2003;  Vujosevic and Krajcinovic, 1997; Saichev and Sornette, 2005].
Many investigations focused on homogeneous materials such as metals
[Ishikawa et al., 2002] and ceramics [Goretta et al., 2001; Morita and Hiraga,
2002] and numerous recent studies are concerned with heterogeneous materials
such as composites and rocks 
[Liu and Ross, 1996; Guarino et al., 2002; Lockner, 1998]. 
Visco-elastic creep and creep-rupture behaviors are among the critical properties needed to
assess long-term survival of material structures.

A given bond linking two neighboring blocks is modeled
as an elastic medium subject to creep and damage, represented
by a non-linear visco-elastic rheology. Following Nechad et al. [2005], the bond is assumed
to be equivalent to an Eyring dashpot in
parallel with a linear spring of stiffness $E$. Its deformation $e$ is governed by
the Eyring dashpot dynamics
\begin{equation}
\label{eyring}
\frac{de}{dt}=K \sinh(\beta s_{\rm dashpot})
\end{equation}
where the stress $s_1$ in the dashpot is given by 
\be
s_{\rm dashpot}={s \over 1-P(e)} -Ee~,
\label{kthnmb}
\ee
Here, $s$ is the total stress applied to the bond and $P(e)$ is the 
damage accumulated within the bond during its history
leading to a cumulative deformation $e$.
$P(e)$ can be equivalently interpreted as the fraction of representative
elements within the bond which have broken, so that the applied stress $s$
is supported by the fraction $1-P(e)$ of unbroken elements.
The Eyring rheology 
(\ref{eyring}) consists, at the microscopic level,
in adapting to the material grain rheology the theory of reaction rates
describing processes activated by crossing potential barriers.
$K$ is a material property and $\beta$ is proportional to the inverse
of an effective temperature of activation that does not need to be 
the thermodynamic temperature [Ciliberto et al., 2001; Saichev and Sornette, 2005].

Following Nechad et al. [2005], we postulate the following dependence
of the damage $P(e)$ on the deformation $e$:
\begin{equation}
P(e)=1- \Bigg( \frac{e_{01}}{e+e_{02 }}\Bigg) ^\xi~,
\label{mgm,tbl;}
\end{equation}
where $e_{01}, e_{02 }$ and $\xi$ are 
constitutive properties of the bond material.
In Nechad et al. [2005], expression~(\ref{mgm,tbl;}) is derived from a mean field
approximation which consists in assuming that the applied load is shared
equally between all representative elements (RE) in the bond: each surviving 
RE is subjected 
to the same stress equal to the total applied force 
divided by the number of surviving RE.
This mean field approximation  has been shown to be a good approximation 
of the elastic load sharing for sufficiently
heterogeneous materials [Roux and Hild,  2002;  Reurings and Alava, 2004].
The exponent $\xi$ is a key parameter, quantifying the degree of 
heterogeneity in the distribution of strengths of the representative elements.
The smaller $\xi$ is, the slower $P(e)$ goes to $1$ as the deformation $e$ increases,
due to the existence of very strong elements still being able to support
the stress. For large $\xi$, $P(e)$ moves rapidly to the value $1$ corresponding to total rupture as soon as $e$ becomes larger than $e_{01}-e_{02 }$.

If the stress remains constant, it is possible to calculate the dynamics of the deformation 
$e(t)$ following from (\ref{eyring},\ref{kthnmb}) and (\ref{mgm,tbl;}). Nechad et al. [2005]
found two regimes depending on the value of $s$ with respect to a reference
value $s^*$, defined as the minimum stress such that there exists at least one
deformation $e$ for which $s_{\rm dashpot}$ given by (\ref{kthnmb}) vanishes:
(i) for $s < s^*$, the bond deformation $e$ converges to a finite asymptotic value
at long times and the bond does not break; (ii) for $s \geq s^*$, the deformation $e$
diverges in a finite time $t_c$ and the corresponding finite-time singularity
in the damage-creep process corresponds to the rupture of the bond.
Appendix B gives a simple formula (\ref{tcs}) for determining $t_c$ for the bond rupture,
when the bond is subjected to a constant stress $s$.

\subsubsection{Integration of stress history}

The analysis of the previous subsection is tractable if the stress
remains constant or its variation with time is simple. In our case, a given
spring is subjected to a series of stress changes associated with the
different sliding events of the two blocks it connects. As a consequence, 
the stress history on a given bond is in general quite complicated.

In order to determine the time at which 
a bond will fail by the accumulation of damage, we need to integrate 
(\ref{eyring},\ref{kthnmb}) and (\ref{mgm,tbl;}) with a variable stress, whose time dependence
reflects the history of sliding of the two adjacent blocks. Generalizing to all the bonds
connecting all the blocks in the system, this leads to unwieldy calculations.
Therefore, when the stress undergoes its first change, we use the following
rheology to take into account arbitrary stress histories: given some
stress history $\sigma(t'), t'\geq0$, a bond is assumed to break 
at some fixed random time, which is distributed according to 
the following cumulative distribution function
\be
F_0(t)\equiv\int_0^t P_0(t')dt'=1-\exp\left\{-\kappa\int_0^t
[\sigma(t')]^{\rho} {\rm d}t'\right\}~.
\label{pdfggh}
\ee
In other words, $F_0(t)$ is the probability that the random time of bond failure
is less than $t$.
This expression amounts to the consideration of a bond failure as a conditional Poisson process
with an intensity which is a function of all the past stress history
weighted by the stress amplification exponent $\rho>0$.
Applied to material failure, this
law captures the physics of failure due to stress corrosion, to
stress-assisted thermal activation and to damage (see Newman et al. [1994; 1995]
and references therein).
The parameter $\rho$ characterizes the material properties of the damage 
process. In our simulations, we will use a typical value $\rho=2$. As shown
by formula (\ref{alphaa}) below, this value 
means that, if the stress is doubled, the waiting time until rupture is divided by four. 
Larger values of $\rho$ would capture a larger sensitivity 
to changes in stress in this history-dependent approach. 

Consider now an element which has a given critical 
rupture time $t_1(s_1)$ given by equation (\ref{tcs}) in Appendix B
associated with an initial stress level $s_1$ which has remained
constant from the inception of the process till present.  Let us assume that the stress 
changes from $s_1$
to  $s_2$ at the time $t<t_1$ due to a change in the
position of the blocks linked by the bond under discussion. As a consequence, 
the critical rupture time is changed from 
$t_1(s_1)$ into a new value $t_{12}(s_2)$, which takes into account that 
some damage has already occurred until time $t$ and that the damage will
continue with a faster (resp. slower) rate for $s_2>s_1$ (resp. $s_2<s_1$).
Newman et al. [1994; 1995] and Saleur et al. [1996]  demonstrated that the rheology
(\ref{pdfggh}) implemented into a hierarchical system of representative elements leads to 
\begin{equation} 
t_{12}(s_2)=t+\alpha(t_1(s_1)-t)
\label{alphaa} 
\end{equation} 
where 
\begin{equation} 
\alpha = (s_2/s_1)^{-\rho}~.
\label{alphaa1} 
\end{equation} 
Note that formula (\ref {alphaa}) with (\ref {alphaa1}) recovers 
the fact that the critical rupture time remains unchanged at $t_1(s_1)$ 
if the stress remains constant at $s_1$.  Note also that, since $\rho>0$, 
$\alpha \leq 1$ which leads to $t_{12}(s_2) < t_1(s_1)$ for $s_2>s_1$, 
as expected.

The new rupture time $t_{12}(s_2)$ (and waiting time $t_{12}(s_2)-t$ to rupture)
can be compared with the rupture time $t_2(s_2)$ that would have been
determined by a constant stress $s_2$ applied from the beginning $t=0$. Indeed,
for constant stresses $s_1$ or $s_2$ applied at $t=0$, the model (\ref {pdfggh})  predicts 
the simple scaling relation
\begin{equation} 
{t_2(s_2) \over t_1(s_1)} = \left( {s_1 \over s_2} \right)^{\rho}  ~.
\label{ghjkbwg}
\end{equation} 
For $s_2 \geq s_1$, expression (\ref{alphaa}) can be written in a form that allows a better
comparison with (\ref{ghjkbwg}) as follows:
\begin{equation} 
{t_{12}(s_2) \over t_1(s_1)} =  {t_2(s_2) \over t_1(s_1)} + 
\left[ 1 - {t_2(s_2) \over t_1(s_1)}\right]  {t \over t_1(s_1)}~.
\label{alphaarga} 
\end{equation} 
The scaling dependence of (\ref{ghjkbwg}) is different from the exponential
dependence predicted by equation (\ref{tcs}), that derives from the model
given in the previous section \ref{tbhothjw}: 
\begin{equation} 
\label{tcshist}
{t_2(s_2) \over t_1(s_1)} = \exp \left( -\gamma (s_2 - s_1) \right)   ~,\;\;~{\rm for}~s_1~\;{\rm and}\;
~s_2~>~s^*~.
\end{equation} 
However, in the limit $\rho \to 0$ and $\gamma \to 0$, both expressions (\ref{ghjkbwg}) 
and (\ref{tcshist}) become equivalent in the leading order of their expansions in powers
of $(s_2-s_1)$. In our simulations, we will use (\ref{alphaa}) as it is the most convenient
to account of all the sliding and rupture events that may occur in the network of blocks
during the damage processes of all bonds. This rule embodies the creep and 
damage processes described in Section \ref{tbhothjw} and by equation
(\ref{pdfggh}).

\subsubsection{Possibility of bond healing}

In addition, we introduce the possibility of bond healing, to account for
self-healing properties of natural material, that may result from sintering and chemical
reactions. Two cases have to be distinguished:

i. In the general case, a block can be disconnected from its neighbors during the
destabilization process. This isolated block will then slide, getting in
contact with and pushing its neighbor situated downwards.
To account for this situation, we assume that these
blocks aggregate. In our numerical procedure, this amounts
to removing the degrees of freedom associated with
this isolated block and to doubling the
mass of the block downward.

ii. To prevent penetration between two blocks, we enable bonds to heal, 
i.e., a broken bond is reintroduced as soon as the distance between two blocks is smaller than a given
threshold.

\subsubsection{Algorithm of the damage process}

In summary, simulation of the damage process leading to bond rupture between
blocks proceeds as follows.
\begin{enumerate}
\item Given an initial configuration of all the blocks within the network, the
elastic forces exerted by all bonds can be calculated from their extension/compression.

\item For each bond $i$ subjected to an initial stress $s_0(i)$, we calculate the corresponding 
critical time $t_{c,0}(i)$ at which it would rupture if neither of the two blocks
connected to it moved in the meantime.
For those bonds where $s_0(i)<s^*$ defined in Equation (\ref{hnnbn}), $t_{c,0}(i)$ is infinite.

\item However, due to the subcritical frictional processes described in Section \ref{frictionaa},
some blocks will move when time reaches their $t_f$ given by (\ref{kthwhlhw}). 
The bond attached to the first unstable block will have its stress modified to
a new value in fast
time during the dynamical sliding event.  Note that the first block to 
slide may trigger an avalanche of block slides, therefore the stress is
modified in more
than just one bond.

\item For all bonds whose stresses have been changed during this ``avalanche,''
we calculate their new critical rupture time using Equation (\ref{alphaa}). In general,
each bond gets a new rupture time which is different from that of other bonds.

\item Some bonds will eventually fail, modifying the balance of forces on their
blocks and accelerating the transition to the sliding regime, after which 
the stresses in the bonds connected to the same blocks are modified.
The modified stresses will again lead to updates in the critical rupture times
for those bonds, according to Expression (\ref{alphaa}).

\item It is important to note that we use a separation of time scale: subcritical frictional
creep and damage by creep preceding rupture occur in ``slow time''
compared with the ``fast time'' dynamics during bond rupture (which is assumed to
occur instantaneously when the critical time $t_c$ of that bond is reached)
and with the sliding dynamics. In other words, we freeze the subcritical frictional
evolution as well as the damage process when blocks are sliding dynamically.
\end{enumerate}

This leads us now to describe the third dynamical regime occurring in our system
of blocks.

\subsection{Dynamical interactions}
\label{dynint}
When a block reaches its critical time $t_f$ for sliding as determined in Section \ref{frictionaa},
it suddenly enters a regime described by Newton's first equation of inertial acceleration:
\begin{equation}
\label{eqmotion}
\sum~\vec F~ =~ m \vec a.
\end{equation} 
Here, the total force $\sum~\vec F$ exerted on a given block
is the sum of the gravitational force, the dynamical frictional resistance discussed 
at the end of Section \ref{frictionaa} and the elastic forces exerted by the four bonds attached
to it. 
\\
The signs and amplitudes of the four forces exerted by the four bonds depend
on the relative position of the block with respect to the positions of its four neighboring blocks.
Denoting by $(x_{i,j}, y_{i,j})$ the coordinates of block $(i,j)$, the total
force exerted by the four springs on this block reads
\begin{equation}
\label{eqFT}
\begin{array}{l}
F_{i,j}^{x} =\\
\;\;\;\;b_{i+1,j}\;E\;\Big[x_{i+1,j}-x_{i,j}-\frac{(x_{i+1,j}-x_{i,j})\;l}{\sqrt{(x_{i+1,j}-x_{i,j})^2+(y_{i+1,j}-y_{i,j})^2}}
\Big]\\
\;\;+b_{i-1,j}\;E\;\Big[x_{i-1,j}-x_{i,j}-\frac{(x_{i-1,j}-x_{i,j})\;l}{\sqrt{(x_{i-1,j}-x_{i,j})^2+(y_{i-1,j}-y_{i,j})^2}}
\Big]\\
\;\;+b_{i,j+1}\;E\;\Big[x_{i,j+1}-x_{i,j}-\frac{(x_{i,j+1}-x_{i,j})\;l}{\sqrt{(x_{i,j+1}-x_{i,j})^2+(y_{i,j+1}-y_{i,j})^2}}
\Big]\\
\;\;+b_{i,j-1}\;E\;\Big[x_{i,j-1}-x_{i,j}-\frac{(x_{i,j-1}-x_{i,j})\;l}{\sqrt{(x_{i,j-1}-x_{i,j})^2+(y_{i,j-1}-y_{i,j})^2}}
\Big]\\
\end{array}
\end{equation}
where $b_{i,j} = 1$ if the bond exists and $b_{i,j} = 0$ if the bond has failed previously
and $E$ is the elastic spring constant.
In terms of symmetry, $F_{i,j}^{y}$ follows by switching $x\leftrightarrow y$.
\\
The dynamical friction force is opposite to the sliding direction and is equal in 
amplitude to $\mu_{\rm dyn}\;P_{\rm weight} \cos \phi$, where $P_{\rm weight}$ is the weight
of the block and $\phi$ is the angle between the surface of contact of the block with the basal plane 
and the horizontal.
\\
To prevent numerical instabilities developing into oscillations 
close to the stopping point of sliding blocks, a viscosity has to be introduced in Equation
(\ref{eqmotion}). This ``numerical'' viscosity has a physical interpretation:
radiated energy coming from vibrations due to jerky block sliding past
asperities has been shown to lead to a pseudo-viscous behavior
[Johansen and Sornette, 1999].
In order to avoid such oscillations, the viscous parameters have to be chosen in such
a way that each block motion is in the overdamped regime.
This leads to a viscous coefficient of :
\begin{equation}
\label{visc}
\eta=\frac{2 \sqrt{m E}}{mg \cos(\phi)}
\end{equation}
This viscous term has to be added to Equation (\ref{eqmotion}).


Note again that  we freeze the subcritical frictional
evolution as well as the damage process when blocks are sliding dynamically
according to Equation (\ref{eqmotion}).
This allows us to decouple the fast dynamical regime described by Equation (\ref{eqmotion})
from the subcritical frictional process occurring
at the interface between each block and the basal surface described in Section \ref{frictionaa} and the damage 
and creep deformation occurring in each bond described in Section \ref{hjjbgmfqq}.

In the general case, as more than one block can slip simultaneously, we will have to solve a
system of dynamical equations like Equation (\ref{eqmotion}), one for each concomitantly
sliding block.

\subsection{Numerical treatment and summary of our model}

Each of the second-order differential equations in (\ref{eqmotion}) is transformed
into a pair of first-order differential
equations. An iteration scheme using the
fourth-order Runge-Kutta algorithm [Press at al., 1994] is adopted to solve each pair of equations.
As explained above, all Equations (\ref{eqmotion}) corresponding to the
sliding blocks are solved
simultaneously. The dynamic slip of one block may trigger the slip of other
blocks, or the rupture of a bond. The slip of each block is computed until
they all stop when their velocity vanishes, and the corresponding ``avalanche'' is
terminated. Then, the subcritical frictional process is refreshed according
to the rules of Section \ref{frictionaa} and the damage due to creep
occurring in each bond is cumulated to its previous history in line with the 
laws given in Section \ref{hjjbgmfqq}.

\paragraph{Summary and global algorithm}

\begin{figure}
\label{algor}
\centerline{
\includegraphics[width=20pc]{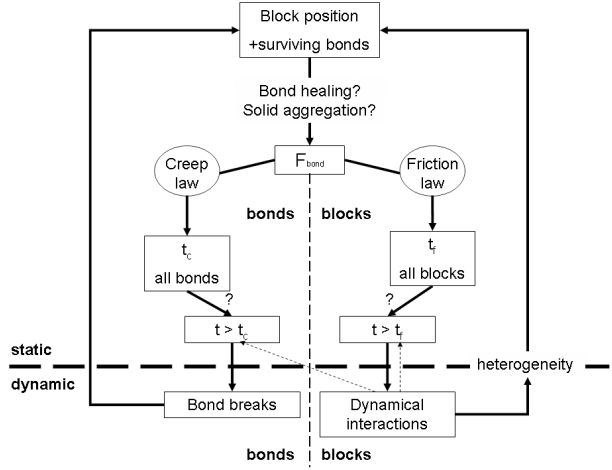}}
\caption{\sf Schematic flowchart of this {\bf modified spring-block} model.
}
\end{figure}

The different steps in this model are described in Figure \ref{algor}. As
explained previously, two phases are distinguished:
\begin{itemize} 
\item[(i)] A quasi-static phase corresponding to the nucleation of block sliding (Section
\ref{frictionaa}) and bond rupture (Section \ref{hjjbgmfqq}).
\item[(ii)] A dynamical phase corresponding to the sliding phase of the blocks (Section \ref{dynint}) 
and of the failure of bonds.
\end{itemize}
To account for the changes of the surface characteristics after blocks have slid,  a
random state parameter $\theta_i$ is assigned to each stopping block.
In this way, the heterogeneity of the basal properties
can be reproduced and is sustained during the evolution of the system.

\section{Classification of the different regimes of gravity-driven instabilities \label{thjetrobj3r}}

In a follow-up paper, we will apply the model presented
in the previous section to case studies of the breaking-off of large ice
masses of hanging glaciers, and in particular to the goal of understanding quantitatively the 
 observations obtained for the Weisshorn hanging glacier [Faillettaz
et al., 2008] and other glaciers in the Swiss
Alps. 

Here, we focus on the general properties of the model. We present
a classification of the possible regimes that emerge
as a result of the interplay between the different mechanisms, whose
amplitudes are controlled by the parameter values.

\subsection{Geometry and parameters}

We consider a system of blocks, whose weights 
remain constant during the numerical simulation.  This
assumption is natural for modeling the nucleation phase
preceding a rockslide or landslide. In the case of glaciers,
this approximation is valid if the thickness of the glaciers
is many times larger than the typical snow accumulation for  a
given year. For instance, in the case of the Weisshorn glacier with
a thickness of about 30 meters, this assumption is well justified
as, in addition, its northeastern face has a steep slope
and is subjected to strong wind, which causes the snow to drift away. 

In order to obtain a realistic description of the damage and fragmentation
that may develop in a sliding mass, we need a sufficiently large number
of blocks. As a compromise between reasonable sampling
and computing time, we use a model composed of $600=20 \times 30$ blocks for
our numerical exploration of the different regimes.

In this vein, consider a 
glacier surface area of approximately $150 \;\rm ~ m^2$, with a depth of 30~m.  
In a model composed of $20 \times 30$ blocks, each block 
corresponds to a discrete mesh of size 30 m thickness, 5
m length and 5~m width. This implies that the weight of each block should be
approximately equal to: $0.7 \times 10^6\;\; \rm kg$ (with a density of $917\; \rm~kg\;m^{-3}$).

We study a particular geometry in which the slope is
divided into two parts: Upper part stable (i.e., slope~$<$~critical slope), 
lower part unstable (slope~$>$~critical slope).
Thus, the array is always driven towards an instability.
Quantitatively, the slope $\phi$ of the underlying supporting surface,
on which the mass is sliding ranges from $\phi_{\rm top}=45^\circ$ (upper part) to $\phi_{\rm bottom}=50^\circ$
(lower part). To account for the curvature of the slope, we used also a
curved elevation of the supporting surface modeled by a portion of a parabola
corresponding to a slope change from  $45^\circ$ to $50^\circ$ over the
size of the length of the glacier. We then chose a static friction parameter
between $\tan 45^\circ$ and $\tan 50^\circ$, so that the upper
part is in a frictional stable state, while the lower part is bound
to start sliding. As a consequence of the slope change, 
the sliding of the lower part will transmit deformation and stress to the upper part.
Our goal is to study this stress and deformation transfer and document
the different regimes of sliding, rupture and fragmentation that occur
as a function of the other model parameters.

The blocks are distributed in a regular mesh along the slope, in such a way
that bonds are initially stress-free. 

The following table summarizes the  parameters used in our simulations
and the values that are fixed throughout our numerical exploration of the
different regimes.
\begin{table*}
\label{creeplawtab}
\begin{center}
\begin{tabular}{|c|c|c|c|c|c|c|c|c|}
\hline
\multicolumn{5}{|c||}{{\bf Geometric parameters}}&\multicolumn{4}{|c|}{{\bf Friction parameters}}\\
\hline
$n_x$&$n_y $& $m_{block}$ & $l$ &\multicolumn{1}{c||}{$\phi$} & A  & $\theta_0$ & $\mu_0$ & \multicolumn{1}{c|}{$\dot{\delta_0}$}\\

- & -& [kg]   & [m]     & \multicolumn{1}{c||}{[$^\circ$]   }      & - &    [d]       & - & \multicolumn{1}{c|}{[$\rm cm.d^{-1}$]} \\

\hline
20& 30 & $2.75 \times 10^6$&10&\multicolumn{1}{c||}{ 40 to 50}&  0.1 &  & 1 &  \multicolumn{1}{c|}{$10^{-3}$ } \\
\hline

\hline

\multicolumn{6}{|c|}{{\bf Creep parameters}} & \multicolumn{3}{c}{} \\ \cline{1-6}
 $E$ & $\beta$ & $C$  & $\xi$ & $e_{01}$ & $e_{02}$& \multicolumn{3}{c}{}\\
 $[\rm Pa.m]$ & [$\rm Pa^{-1}$] & [s]& -     & -   &   -  & \multicolumn{3}{c}{}\\ \cline{1-6}
 $ $  & $ $ & $ $ & 10 & 0.003&0.003& \multicolumn{3}{c}{} \\\cline{1-6}
\end{tabular}
\end{center}
\caption{Parameters used for the simulation. The mass 
has a total number of $n_x \times n_y$ blocks. The second row gives the list of parameters. The third row 
specifies the units of the parameters and the fourth row gives the numerical values used
in the simulation for those parameters which have been fixed throughout. Missing entries correspond
to the parameters that have been varied as explained in the text.}
\end{table*}

We find that the dependence of the mass evolution as a function of the other parameters
can be reduced to a set of two reduced parameters.
\begin{enumerate}
\item The first important parameter is the elastic coefficient $E$ controlling 
the rigidity of the springs transferring stress between the blocks.
\item The second key parameter is the ratio $T_c/T_f$ of two characteristic time
scales associated with the two fundamental processes: internal damage/creep
and frictional sliding. The first time scale $T_c$ is associated with the creep-damage process
occurring in the springs linking the blocks, also accounting for the natural
frequency of the springs 
and reads
\begin{equation}
\label{ctc}
T_{c}=T \;\frac{\ln(\frac{C}{T})}{\beta~s^{\star}}~,
\end{equation}
where $T$ is the natural period of the spring-mass system equal to $2 \pi
\sqrt{\frac{m_{block}}{E}}$, and where $C$, $\beta$ and $s^{\star}$ are three parameters of the creep-damage law defined
by expressions (\ref{tkhtnh2t}),  (\ref{geneyring}) and (\ref{hnnbn}). 
Note that 
$\gamma$ in expression (\ref{hy3ghte}) reduces to $\beta$ since 
we assume for simplicity that $ e_{02}=e_{01}$.\\
The second time scale $T_f$ is associated with the frictional process
controlling the tendency of blocks to slide and is derived
from expression (\ref{tkhntrk}), leading to
\begin{equation}
\label{ctf}
T_f=\frac{\theta_0}{\exp(\frac{\tan(\phi_{\rm top})-\tan(\phi_{\rm bottom})}{A})-1}~.
\end{equation}
Here,  $\theta_0$ and $A$ are two parameters of the friction law and 
$\phi_{\rm top}$ and $\tan(\phi_{\rm bottom})$ are the two angles
of the unevenly slopping basal plane supporting the instable
sliding mass.
\end{enumerate}
Note that the first parameter $E$ has also an impact on $T_{c}$ through its
effect on $s^{\star}$.

\subsection{The three main regimes}

\subsubsection{Coherent sliding of the whole mass: $T_c/T_f$ large \label{thgtnbvw}}

\begin{figure*}[t]
\begin{tabular}{lr}
\hspace{-2cm}
\begin{minipage}{20pc}
\centerline{
\includegraphics[width=20pc]{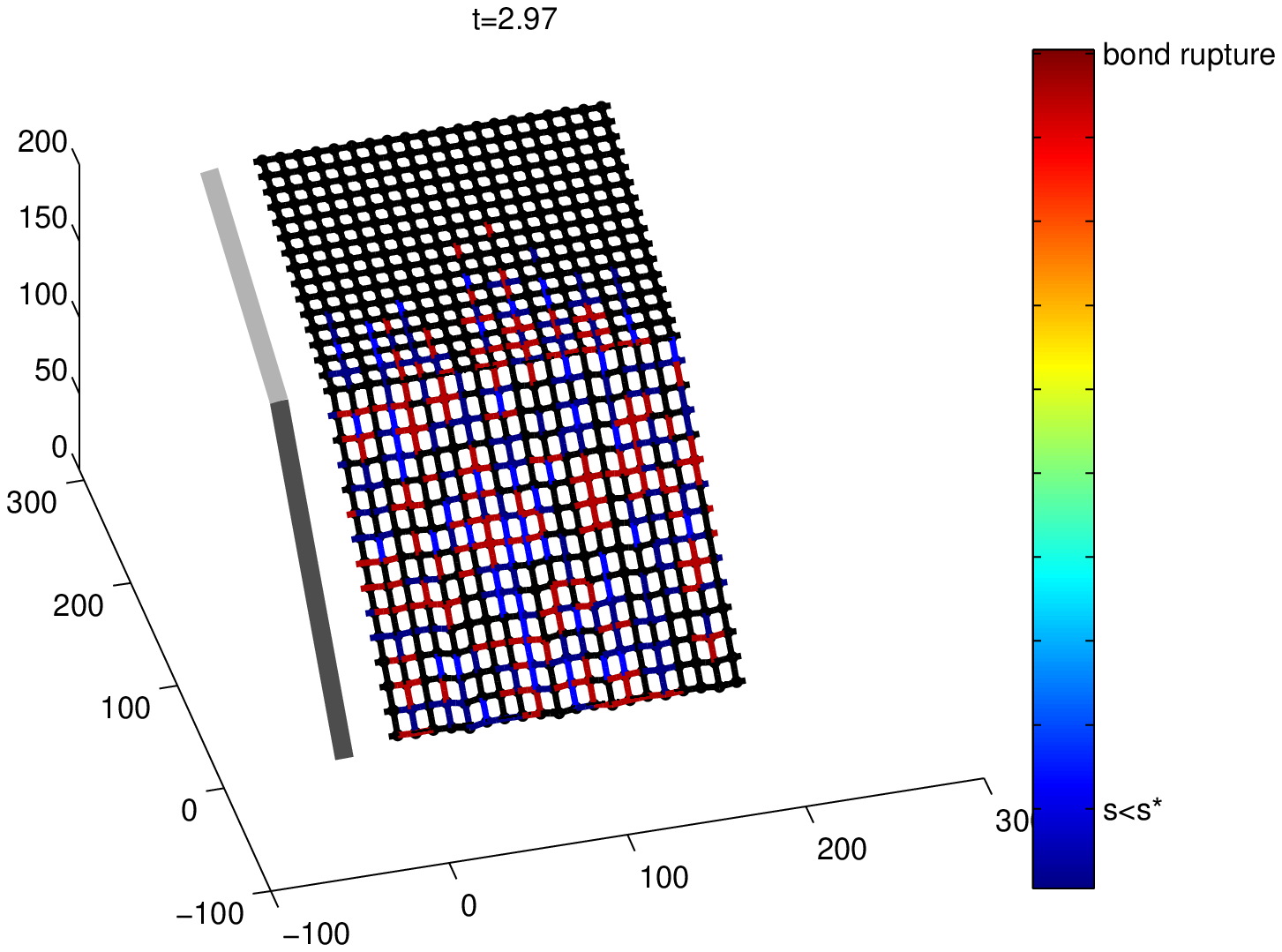}}

\end{minipage}
&
\hspace{-1cm}
\begin{minipage}{20pc}

\centerline{
\includegraphics[width=20pc]{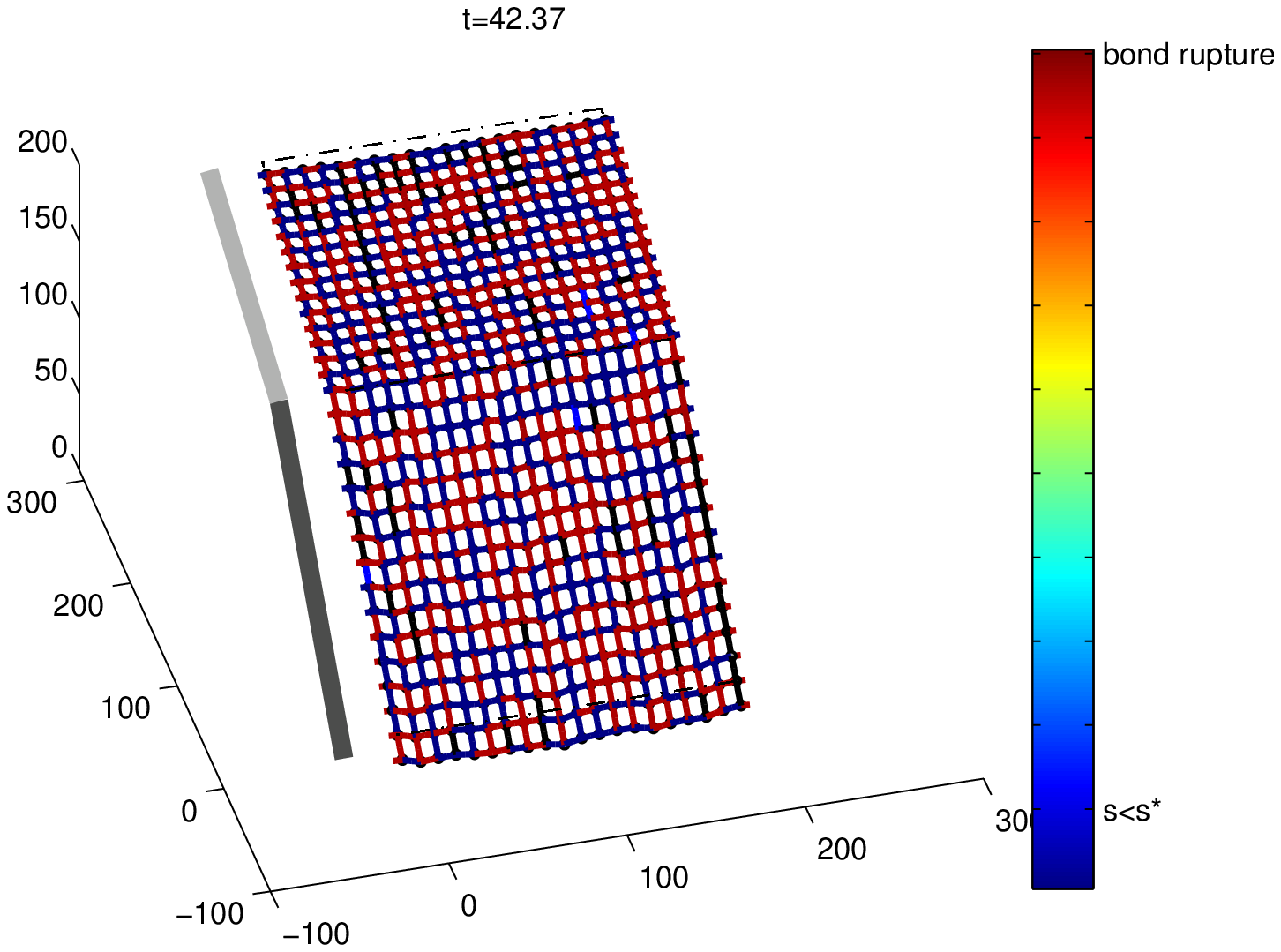}}

\end{minipage}
\\
\hspace{-2cm}
\begin{minipage}{20pc}
\centerline{
\includegraphics[width=20pc]{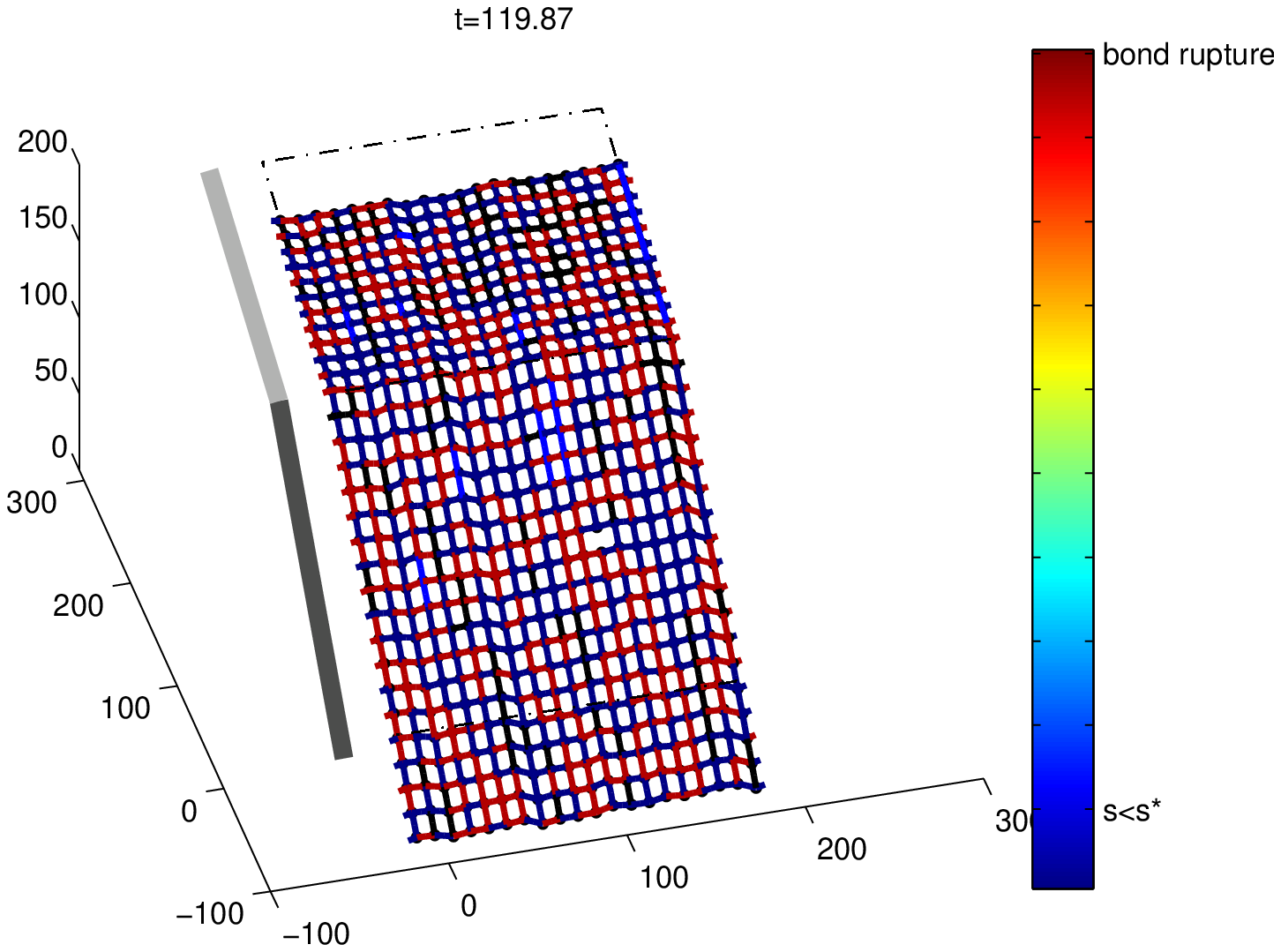}}

\end{minipage}
&
\hspace{-1cm}
\begin{minipage}{20pc}

\centerline{
\includegraphics[width=20pc]{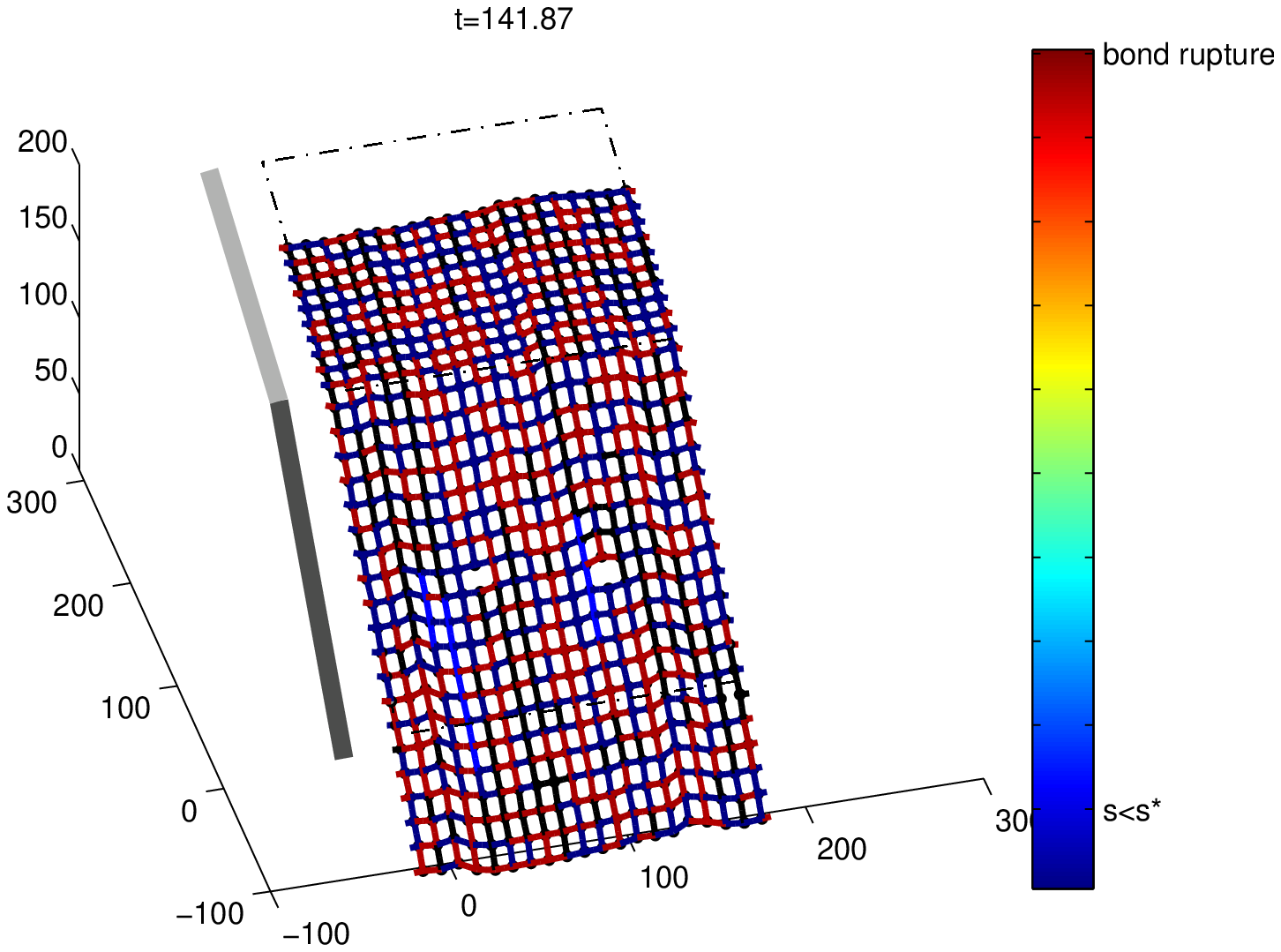}}

\end{minipage}
\label{stick}
\end{tabular}
\caption{Illustration of the propagation of stick-slip events
within the whole mass until it starts to slide coherently 
in a catastrophic runaway, occurring for large $T_c/T_f$ (i.e. $\sim 386$), for any $E$. Time increases from left to right
and top to bottom. 
For the sake of simplicity, block are not drawn. The color of a bond evolves from black
to red as the stress changes from lower than $s^\star$ to close to rupture.
}
\end{figure*}

In this regime, the time needed for damage initiation is so great that the whole
mass undergoes a series of internal stick and slip events, associated with an initial
slow average downward motion of the whole mass. This average motion of the mass
progressively accelerates until a global coherent run-away is observed. Specifically, 
we observe the following evolution:
\begin{enumerate}
\item[(i)] First, the blocks in the lower part, where the slope is larger, start sliding 
intermittently in a stick-slip fashion. 
\item[(ii)] The motion of these blocks transfers stress to the adjacent
blocks, in turn leading to sliding events.
\item[(iii)] This dynamic evolves upstream, leading to an increasing number of
blocks sliding in a stick-slip fashion.
\item[(iv)] The whole system is then undergoing stick-slip sliding,
with increasing synchronization between the sliding blocks.
\item[(v)] With increasing time, the system reaches a regime where
a sufficiently large fraction of blocks are in the unstable lower part
and the mass starts to slide as a whole, leading to a 
catastrophic coherent global slide.
\end{enumerate}
This sequence is illustrated in Figure \ref{stick}.
This regime is observed at smaller values of $T_c/T_f$ for small elastic coefficient $E$.
It is observed for large values of $E$ only for much larger ratios $T_c/T_f$
as shown in the phase diagram of Figure \ref{phasediag}.

\begin{figure*}[h]
\begin{tabular}{lr}
\hspace{-2cm}
\begin{minipage}{20pc}
\centerline{
\noindent\includegraphics[width=20pc]{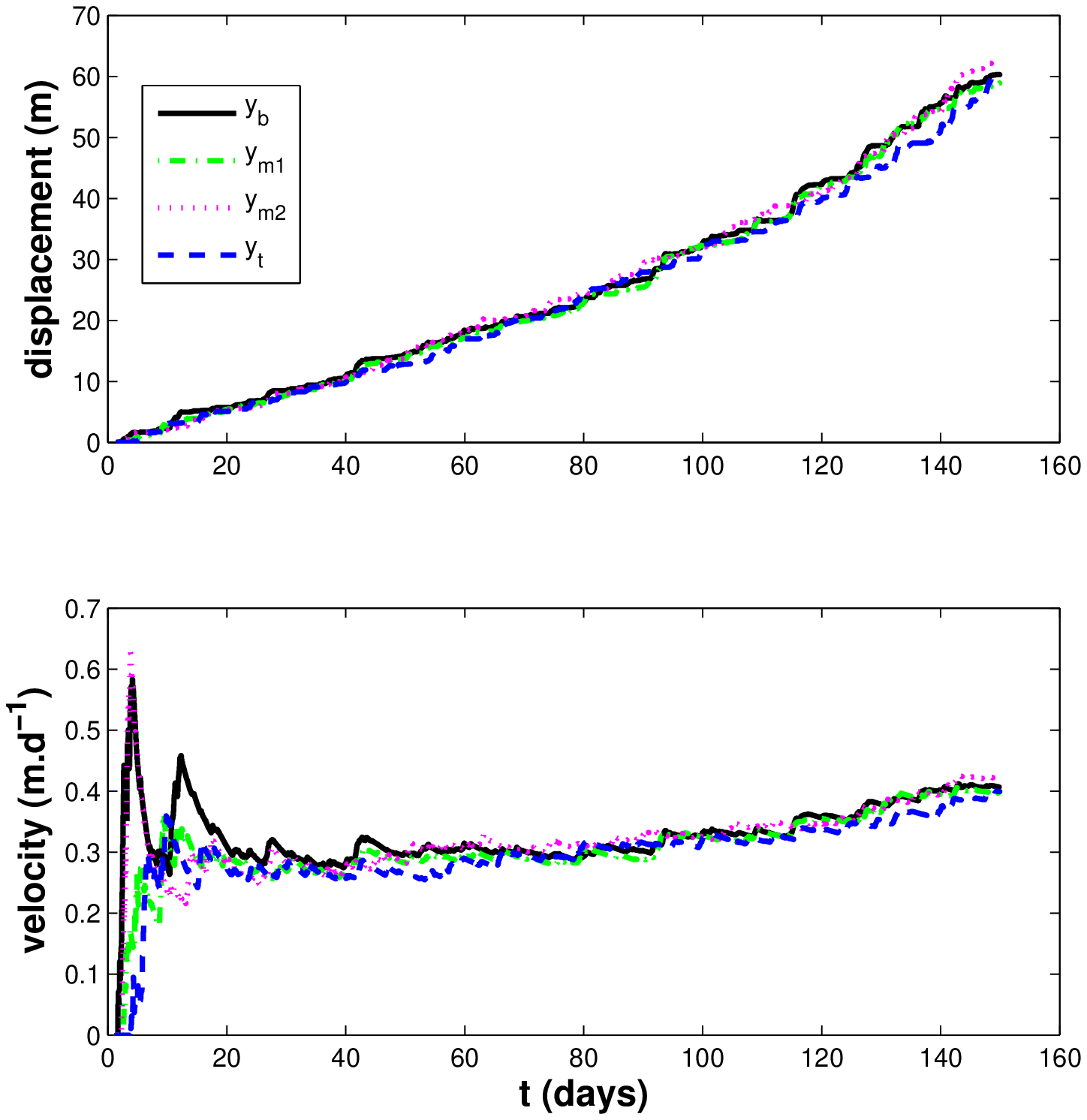}}
\caption{\label{disp_stick} Displacements and velocities of four selected blocks, one taken at the top of
 the slope ($y_t$), two near slope change ($y_{m1}$ and $y_{m2}$) and one at the
 bottom ($y_t$), in the stick-and-slip regime.
}

\end{minipage}
&
\begin{minipage}{20pc}

\centerline{
\noindent\includegraphics[width=20pc]{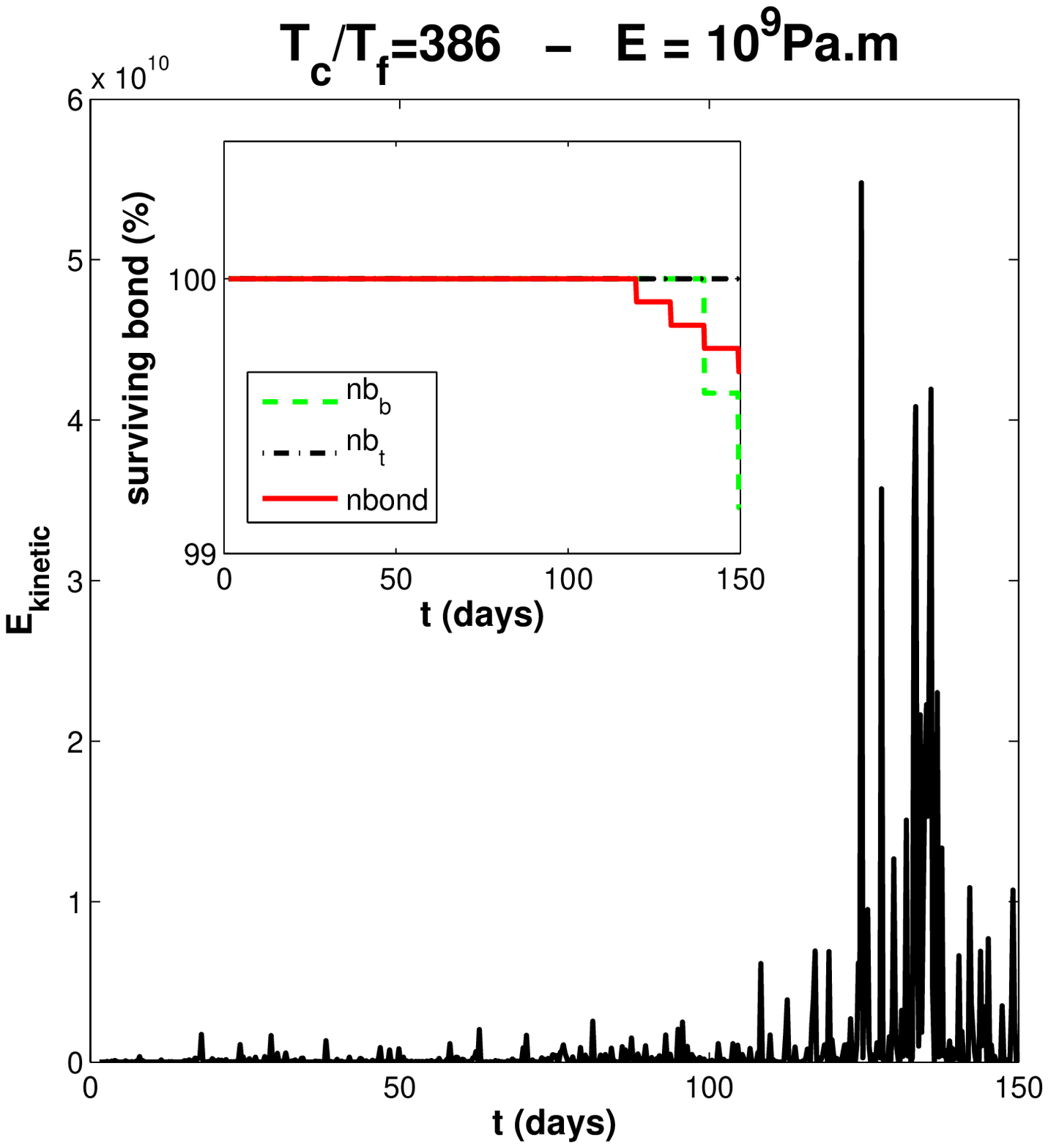}}
\caption{Kinetic energy for the stick-and-slip regime. The total number of surviving
  bonds is shown in the inset (solid line), as well as the number of surviving
  bonds in the upper (dashed and dotted
  line) and lower part (dashed line) of the model.
\label{Ec_stick}}
\end{minipage}
\end{tabular}

\end{figure*}

Figure \ref{disp_stick} shows the time-dependence of four selected blocks, one
in the upper region, two near the slope change and one in the lower region. 
After an initial phase, the velocity stabilizes and slightly increases as the whole
mass slides downwards, where the slope is greater.

The inset in Figure \ref{Ec_stick} shows the number
of surviving bonds as a function of time in the upper and lower
parts. One observes that damage accumulates in the lower
unstable sliding region, whereas the upper part remains basically undamaged.
Figure \ref{Ec_stick} shows the kinetic
energy of the moving blocks as a function of time. This graph gives an
indication of the radiation energy
emitted by cracks forming as a consequence of damage
accumulation. In this regime, in which the whole mass evolves
progressively towards a global coherent run-away sliding, one
can observe a rich precursory behavior long before this
unstable run-away occurs.


\subsubsection{Fragmentation: $T_c/T_f$ small}

\begin{figure*}[t]
\begin{tabular}{lr}
\hspace{-2cm}
\begin{minipage}{20pc}
\centerline{
\includegraphics[width=20pc]{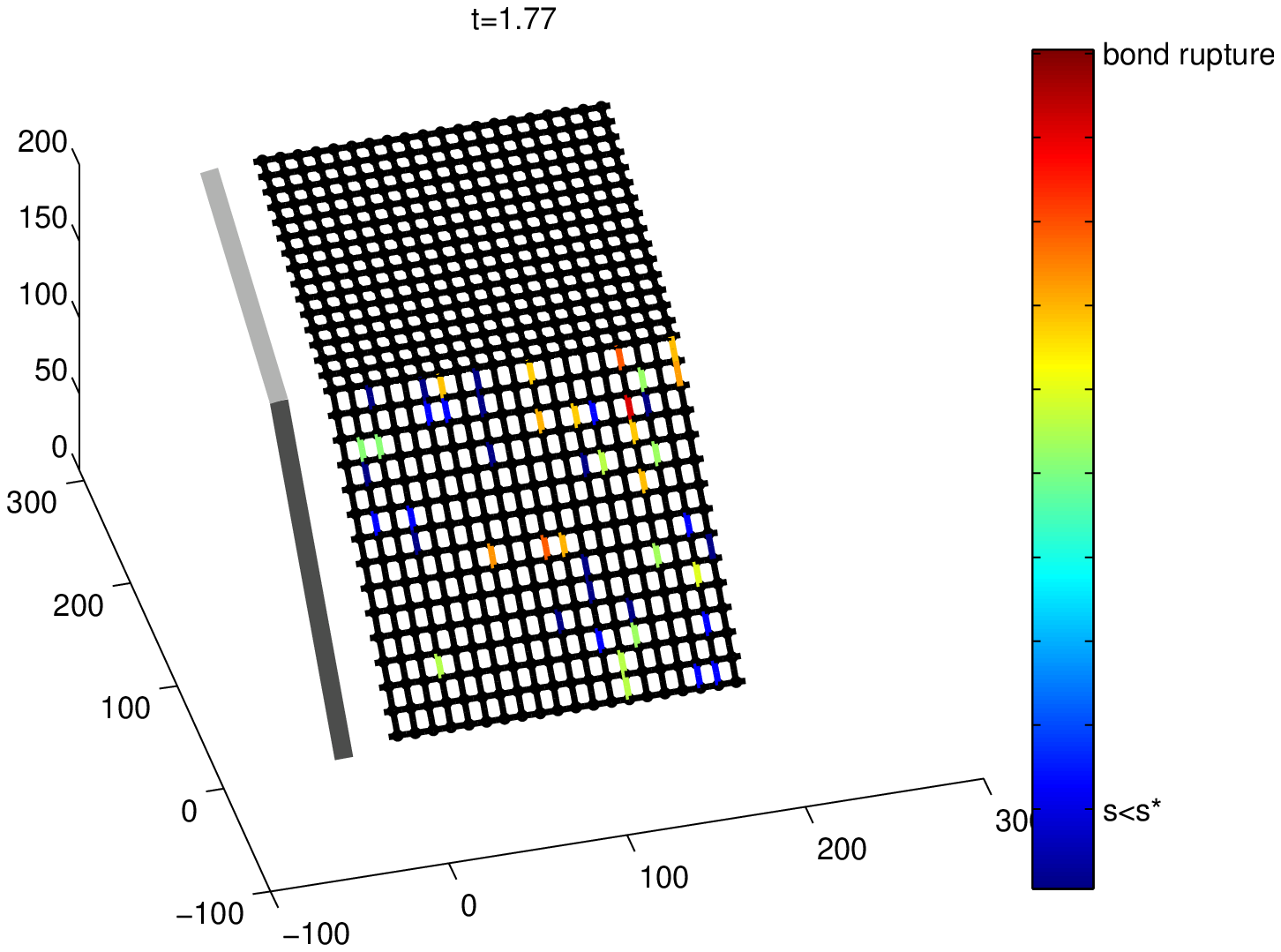}}

\end{minipage}
&
\hspace{-1cm}
\begin{minipage}{20pc}

\centerline{
\includegraphics[width=20pc]{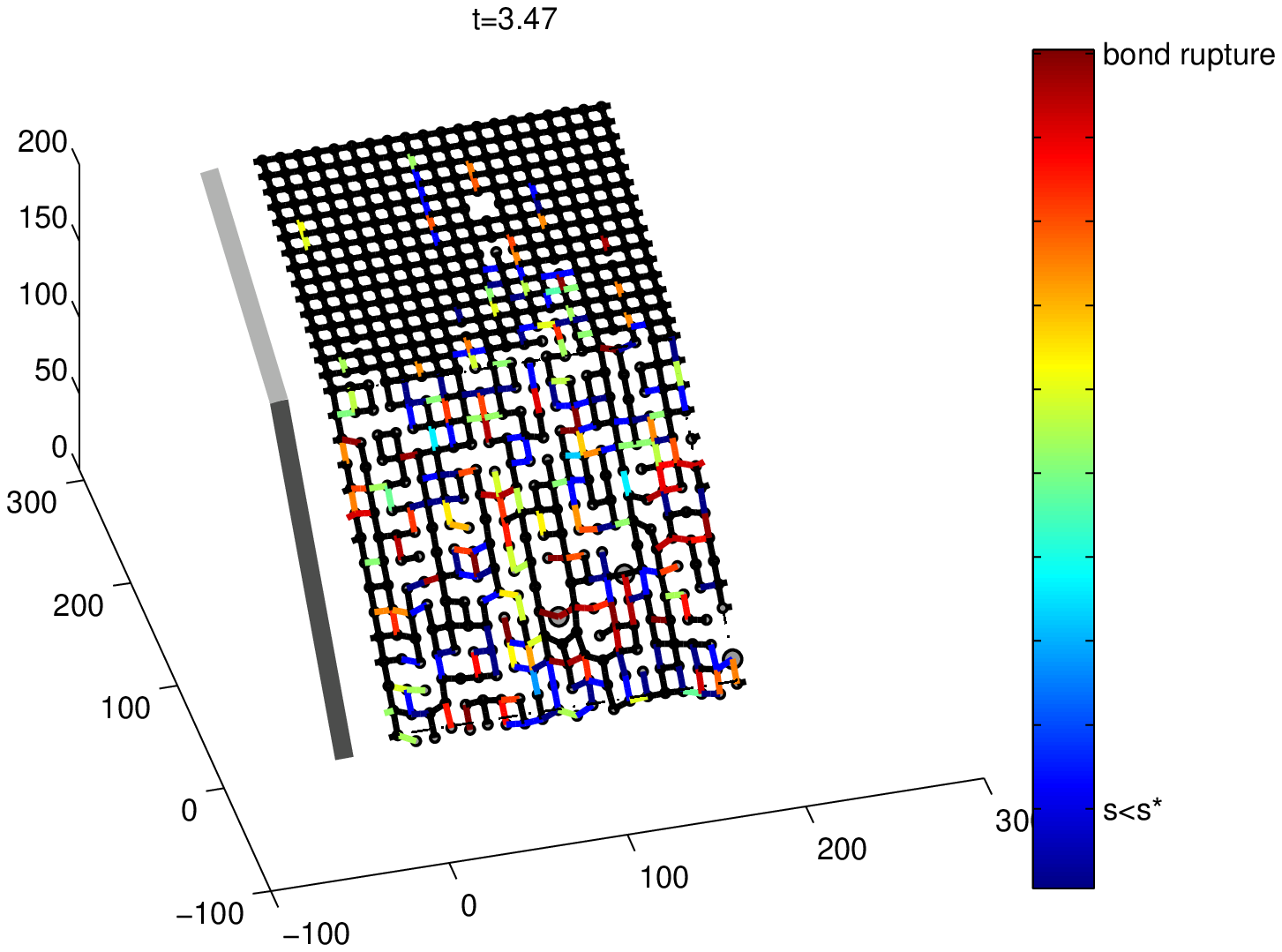}}

\end{minipage}
\\
\hspace{-2cm}
\begin{minipage}{20pc}
\centerline{
\includegraphics[width=20pc]{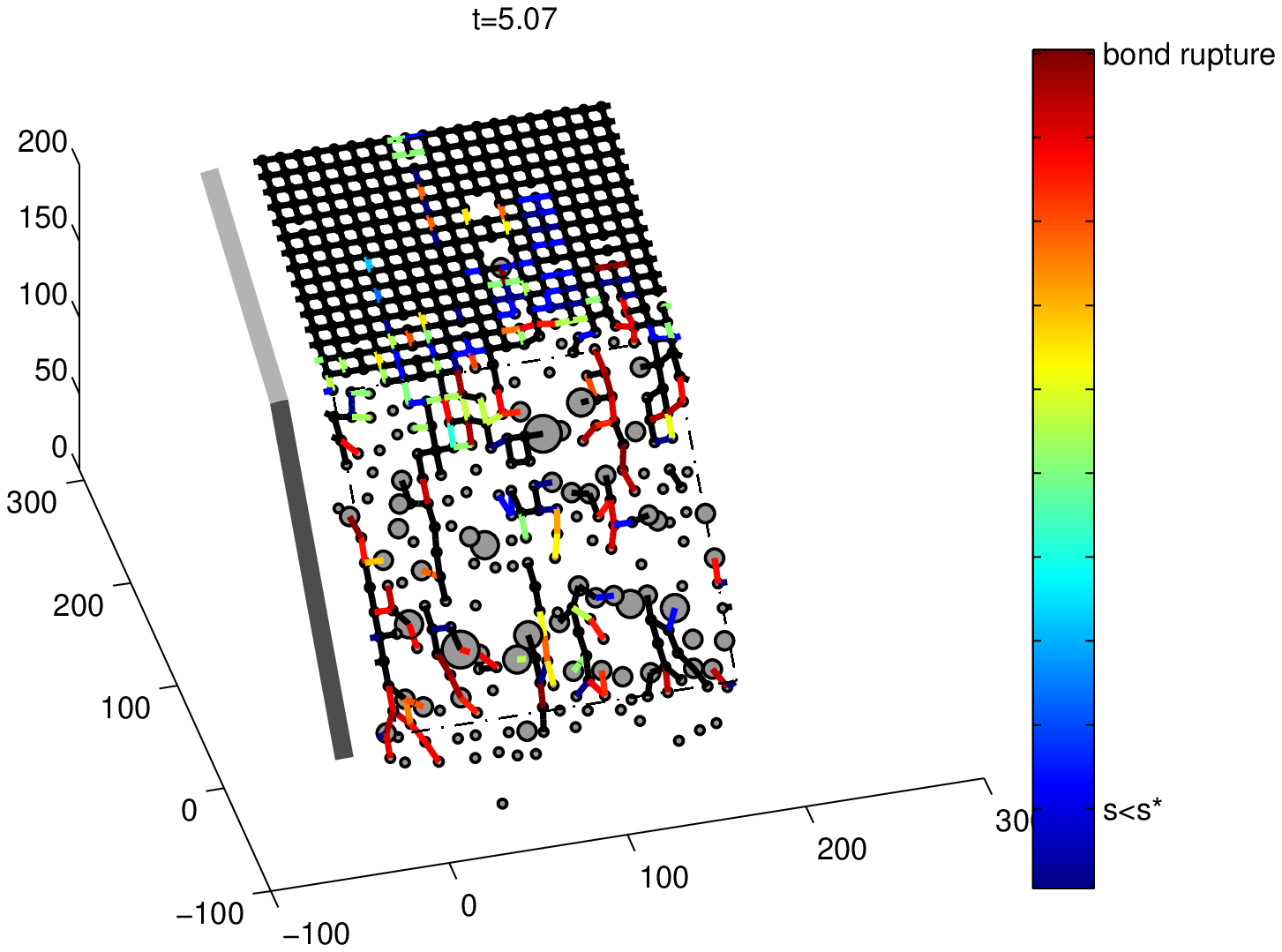}}

\end{minipage}
&
\hspace{-1cm}
\begin{minipage}{20pc}

\centerline{
\includegraphics[width=20pc]{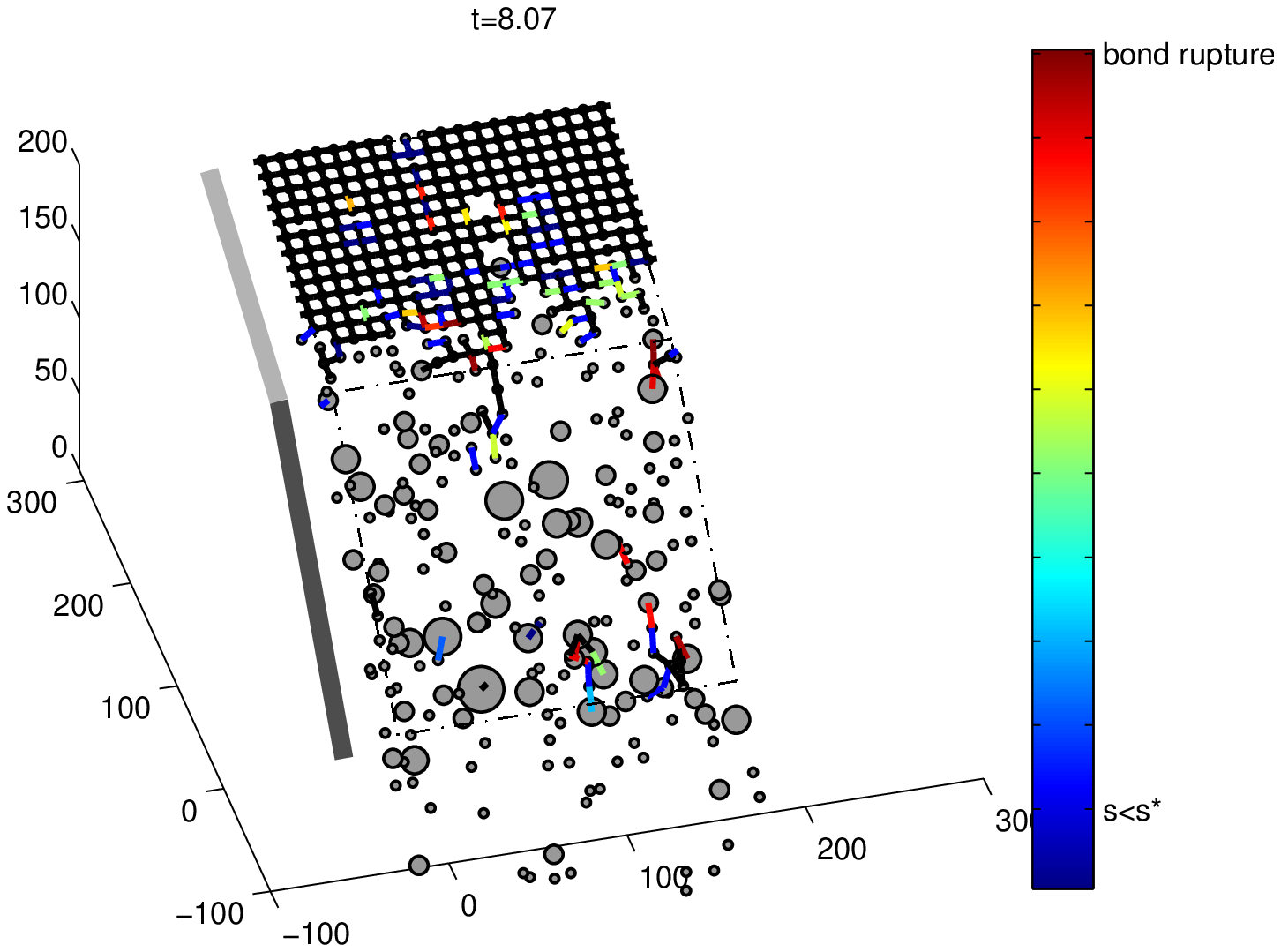}}

\end{minipage}

\end{tabular}
\caption{Illustration of the fragmentation process associated with sliding,
occurring for small $T_c/T_f$(i.e. $\sim 1$), for any $E$. Diameters of the gray circles are
proportional to the corresponding block mass (that change after aggregation process).
}
\label{fragm}
\end{figure*}

This regime has characteristics which are in a sense opposite to those
of the previous regime: creep/damage occurs
before the nucleation of sliding has time to develop, so that bonds break and the bottom part of the mass
undergoes a fragmentation process with the creation of a heterogeneous population of 
sliding blocks.  The typical evolution of the unstable mass is as follows:
\begin{enumerate}
\item[(i)] First, the blocks in the lower part start sliding as in the previous case. 
\item[(ii)] As previously, the deformation due to these sliding events 
propagates upstream, leading to stick-slip events.
\item[(iii)] Due to the relatively fast damage process, springs between
blocks break, 
further increasing the
stress, promoting further spring rupture, and so on.
\item[(iv)] Large fractured zones appear in the lower part,
and blocks become isolated.
\item[(iv)] The final state is characterized by detached isolated blocks.
\end{enumerate}
Four snapshots of the evolution of the mass illustrate this
fragmentation process in Figure \ref{fragm}.
This regime is observed for small values of $T_c/T_f$ and the large elastic
coefficient $E$.
It is observed for small values of $E$, ranging up to much larger ratios $T_c/T_f$,
as shown in the phase diagram of Figure \ref{phasediag}.

\begin{figure*}[h]
\begin{tabular}{lr}
\hspace{-2cm}
\begin{minipage}{20pc}
\centerline{
\noindent\includegraphics[width=20pc]{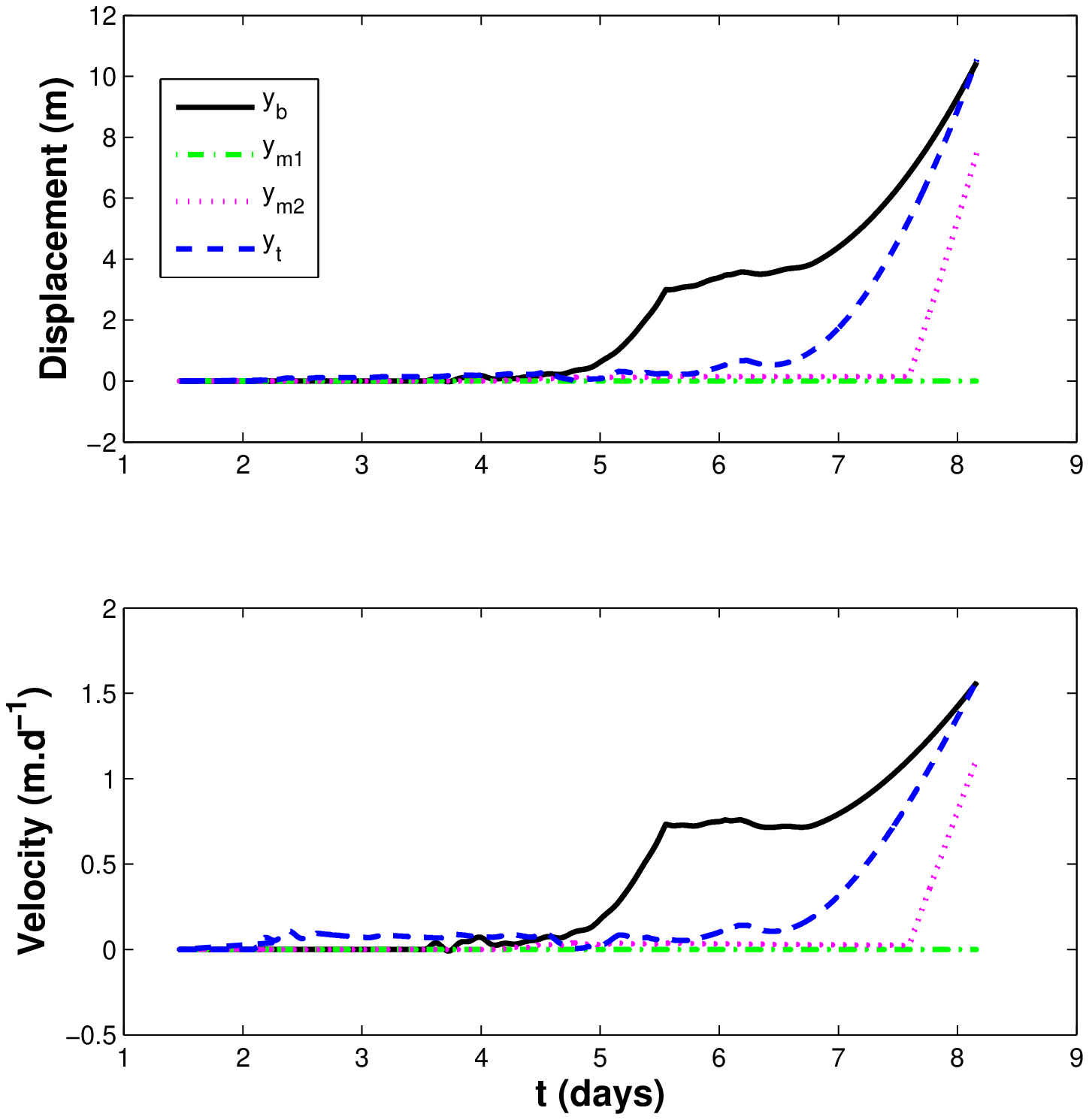}}
\caption{\label{disp_fragm} Displacements and velocities of four selected blocks, one taken at the top of
 the slope ($y_t$), two near slope change ($y_{m1}$ and $y_{m2}$) and one at the
 bottom ($y_t$), in the fragmentation regime.
}

\end{minipage}
&
\begin{minipage}{20pc}


\centerline{
\noindent\includegraphics[width=20pc]{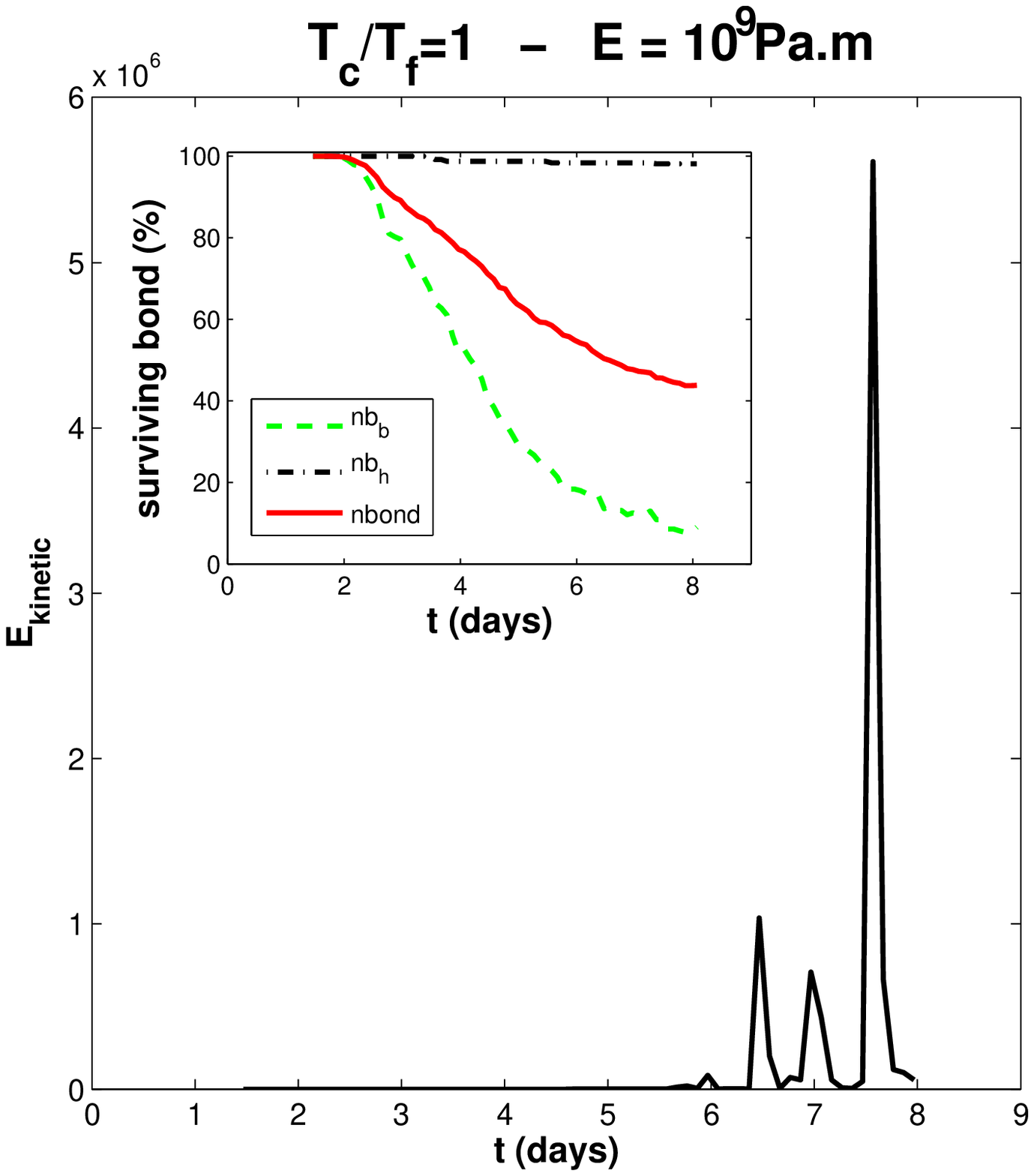}}
\caption{Kinetic energy in the fragmentation regime. The total number of surviving
  bonds is shown in the inset (solid line), as well as the number of surviving
  bonds in the upper (dashed and dotted
  line) and lower part (dashed line) of the model.
}
\label{Ec_fragm}
\end{minipage}
\end{tabular}
\end{figure*}

Figure \ref{disp_fragm} shows the displacement and velocity of four selected
blocks (same as Figure \ref{disp_stick}). Note that the blocks situated in the
upper part do not slide. As soon
as the blocks start sliding, their velocity increases. Generally speaking, this can
be explained by  bond rupture (as $T_c/T_f$ is small),
meaning that the block is alone and does not interact anymore with its neighbors. The
block accelerates according to equation (\ref{eqmotion}) with F=0.

The figure \ref{Ec_fragm} inset shows the time-dependence
of the surviving bonds in the upper and lower
parts. 
A very substantial level of damage can be observed, both
in the lower and upper parts, in contrast with the previous regime. The damage level
in the lower part leads to a completely fragmented mass.
Figure \ref{Ec_fragm} plots the time-dependence of the kinetic
energy of the moving blocks. One can observe
an intermittent kinetic energy that increases prior to  
full fragmentation, thus providing, as
in the previous regime, an unambiguous precursory warning, albeit with
a smaller advance time compared with the previous regime.

The kinetic energy increases due to isolated blocks moving downwards after
the rupture of their bonds (this increase starts at t=7, see Figure
 \ref{fragm}). Because the system is already fragmented, this increase is of
 no use for an early warning. 
Radiated energy seems to provide a more reliable indication of early destabilization
of the system.

\subsubsection{Slab avalanche:  $T_c/T_f \sim 1$}

\begin{figure*}[t]
\begin{tabular}{cc}
\hspace{-2cm}
\begin{minipage}{20pc}
\centerline{
\includegraphics[width=20pc]{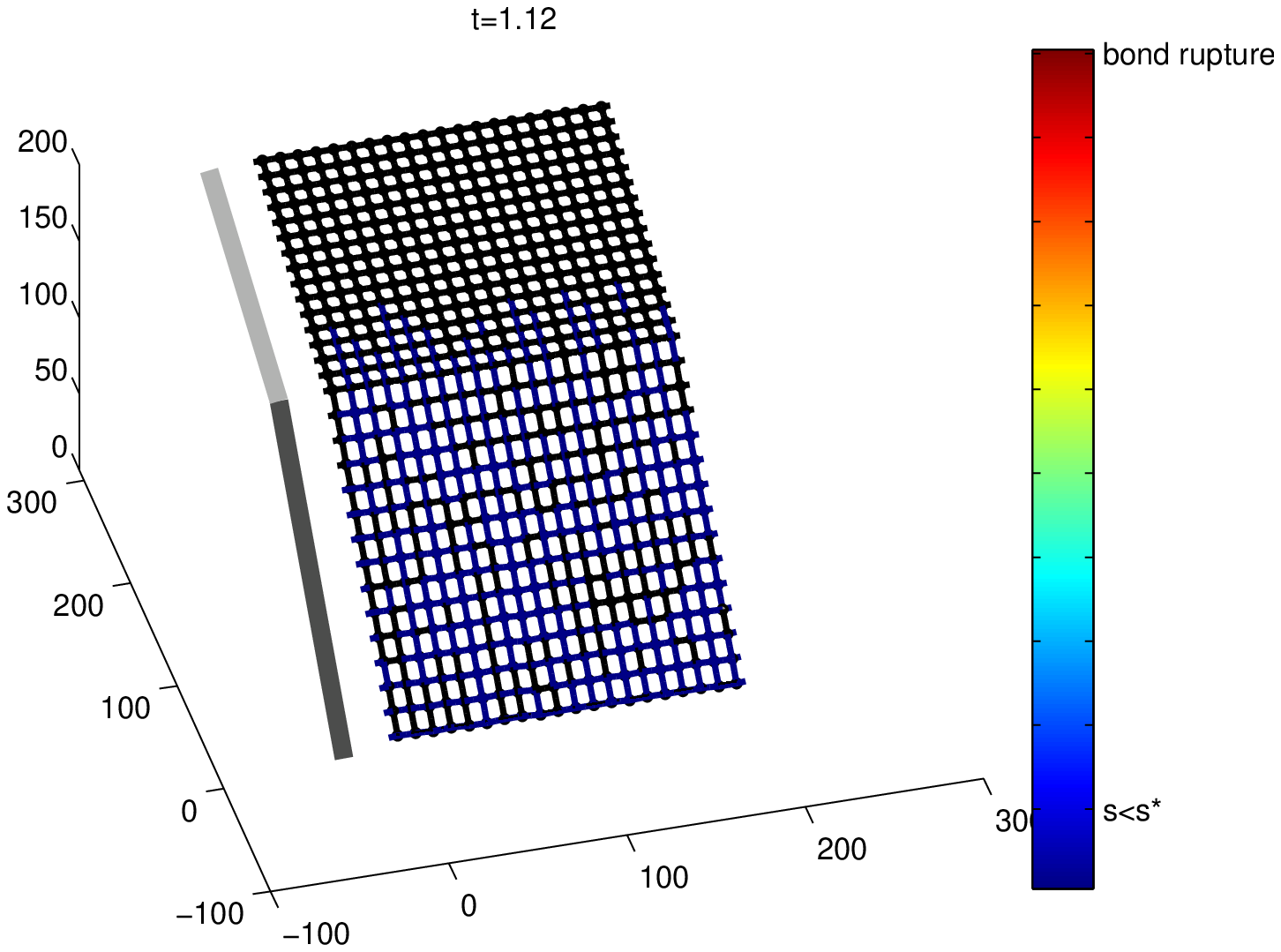}}

\end{minipage}
&
\hspace{-1cm}
\begin{minipage}{20pc}

\centerline{
\includegraphics[width=20pc]{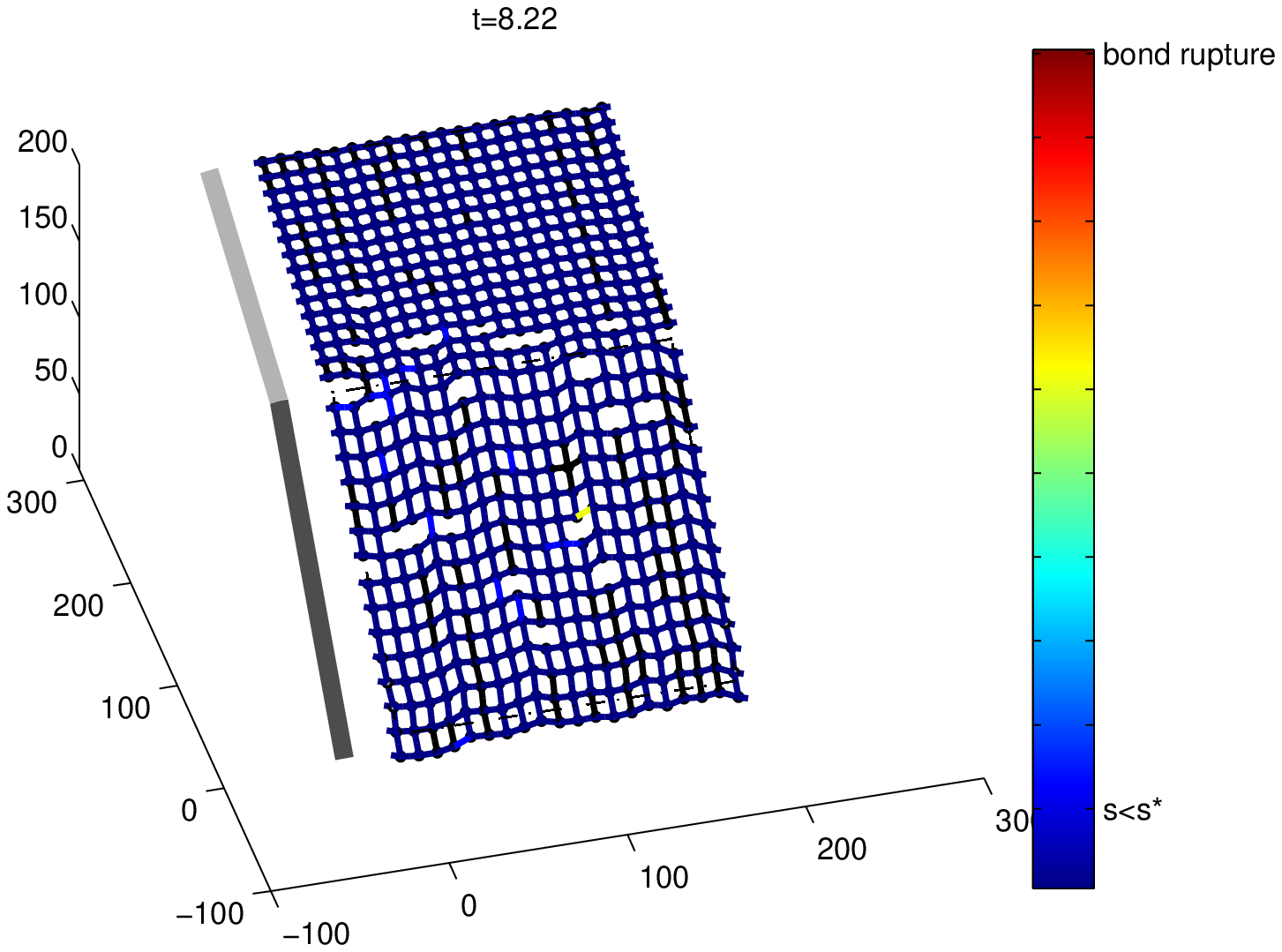}}

\end{minipage}
\\
\hspace{-2cm}
\begin{minipage}{20pc}
\centerline{
\includegraphics[width=20pc]{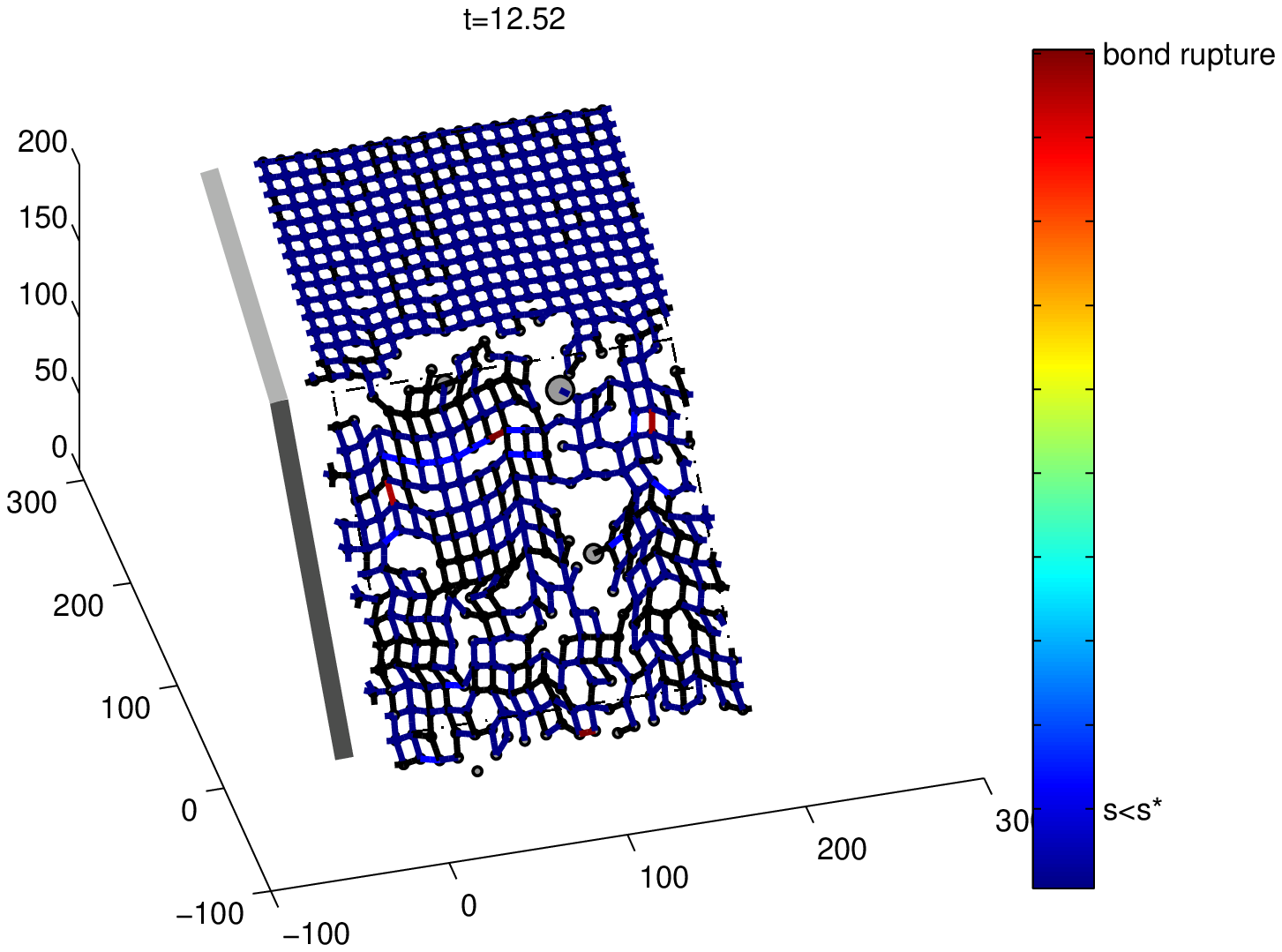}}

\end{minipage}
&
\hspace{-1cm}
\begin{minipage}{20pc}

\centerline{
\includegraphics[width=20pc]{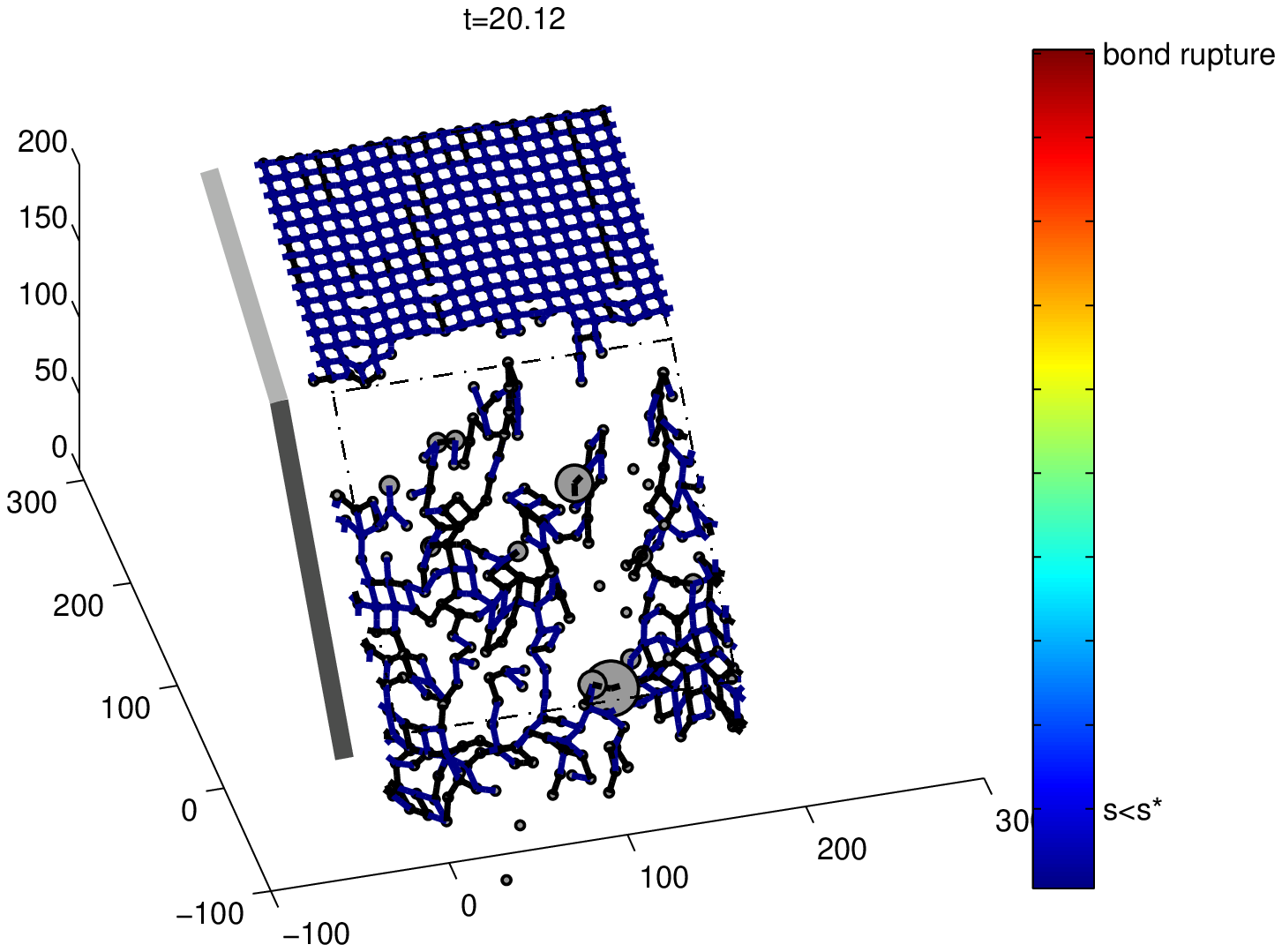}}

\end{minipage}

\end{tabular}
\caption{Illustration of slab avalanche process associated with sliding,
occurring for intermediate $T_c/T_f$ (i.e. $\sim 25$). Diameters of the gray circles are
proportional to the corresponding block mass (that change after aggregation process).
}
\label{slab}
\end{figure*}

When neither damage nor frictional sliding dominates, an interesting
phenomenon is observed: the occurrence of a macroscopic crack
propagating roughly along the location of the largest curvature
associated with the change of slope from the stable frictional state
in the upper part to the unstable frictional sliding state in the lower part.
The evolution of the mass proceeds as follows:
\begin{enumerate}
\item[(i)] The first steps are similar to those observed in the two
previous regimes, with individual blocks starting a stick-slip motion intermittently
in the lower part. 
\item[(ii)] The internal stresses created by these
stick-slip events lead to an upstream propagation of these
events, which progressively invades all the blocks in the lower part
of the mass lying on the largest slope.
\item[(iii)] The damage time scale is such that the boost provided
by the increased stress leads to a concentration of cracking at the boundary
of the two slopes where the curvature is largest.
\item[(iv)] This leads to a nucleation and growth of a macroscopic crack
which separates the mass into two parts. 
\item[(v)] The lower part evolves in a run-away situation with synchronized
accelerating sliding, as in the first regime described in Section \ref{thgtnbvw}.
The upper part remains largely untouched and stable.
\end{enumerate}
This regime is illustrated in four snapshots shown in Figure \ref{slab}.
It occurs in a band of large values of $T_c/T_f $ ratios, for the small enough elastic coefficient $E$,
as shown in the phase diagram of Figure \ref{phasediag}.

\begin{center}
\begin{figure*}[t]
\begin{tabular}{rl}
\hspace{-2cm}
\begin{minipage}{20pc}
\centerline{
\noindent\includegraphics[width=19pc]{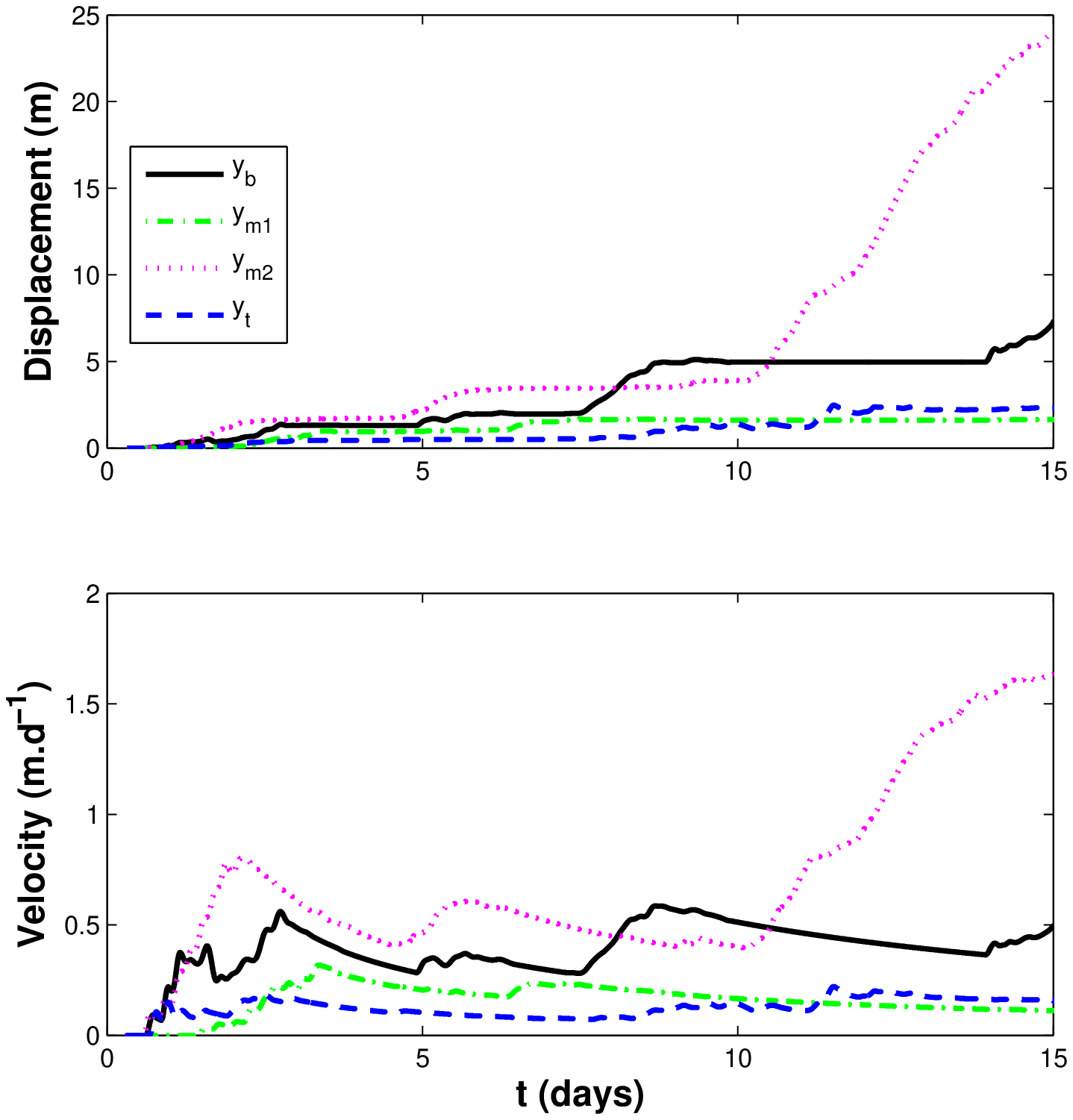}}
\caption{Displacements and velocities of four selected blocks, one taken at the top of
 the slope ($y_t$), two near slope change ($y_{m1}$ and $y_{m2}$) and one at the
 bottom ($y_t$), in the slab avalanche transition regime.
}
\label{disp_slab}
\end{minipage}
&
\begin{minipage}{20pc}
\centerline{
\noindent\includegraphics[width=19pc]{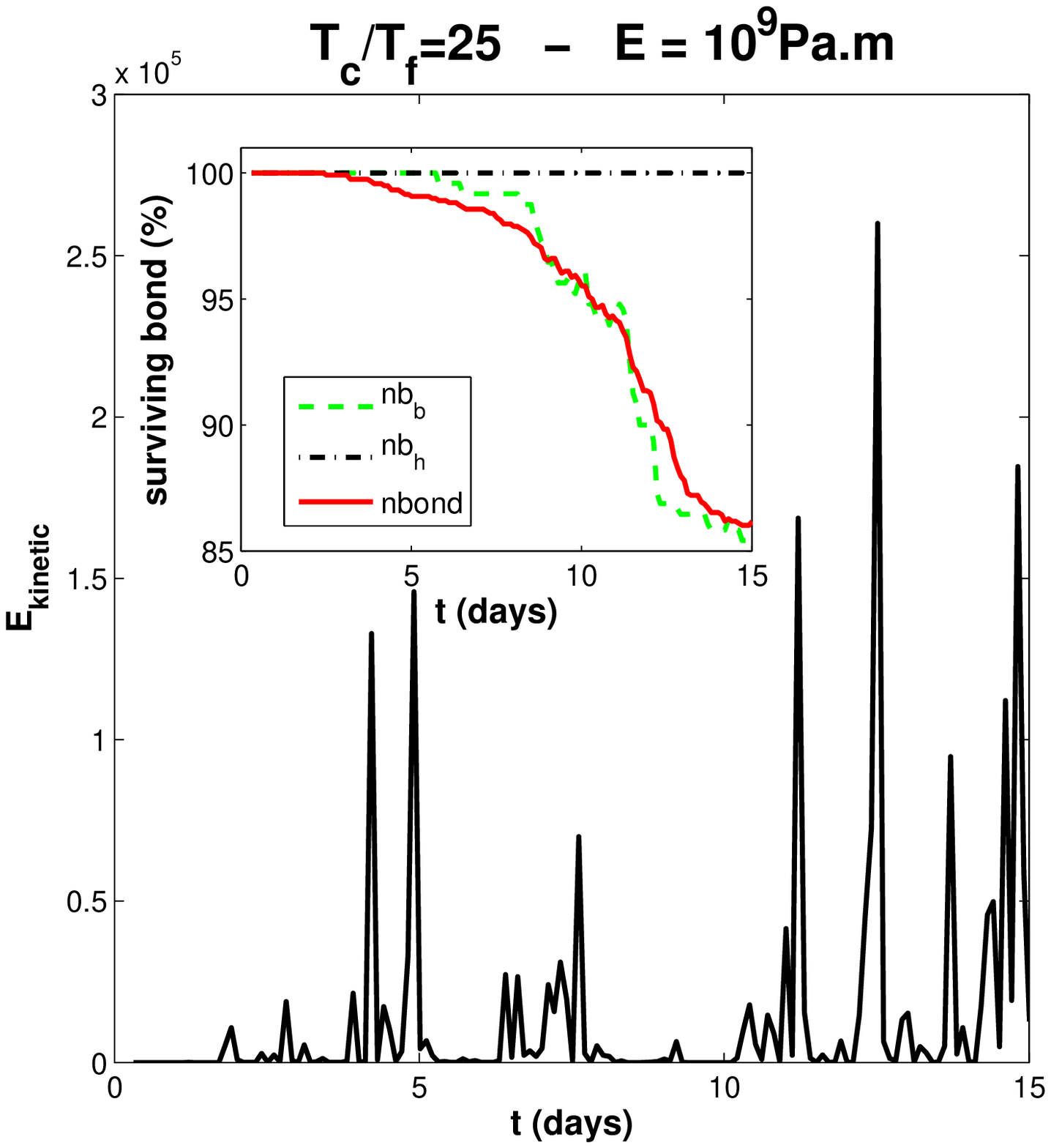}}
\caption{Kinetic energy, in the slab avalanche transition regime. The total number of surviving
  bonds is shown in the inset (solid line), as well as the number of surviving
  bonds in the upper (dashed and dotted
  line) and lower part (dashed line) of the model.
}
\label{Ec_slab}
\end{minipage}

\end{tabular}
\end{figure*}
\end{center}

The Figure \ref{Ec_slab} inset shows the number of surviving bonds as a function of time in the upper and lower
part. Most of the damage
occurs in the lower unstable sliding region. It can be observed that the
fraction of surviving bonds drops suddenly after the nucleation time (here about $13$ days)
of the macroscopic crack separating the lower from the upper parts.
Figure \ref{Ec_slab} shows the time evolution of the kinetic
energy of the moving blocks. Compared with
previous regimes, there is a much shorter (if any) advance warning time,
which is compatible with the first-order nature of the nucleation
of a macroscopic crack. 

\subsection{Phase diagram in the plane $[E; T_c/T_f]$}

\begin{figure}
\centerline{
\includegraphics[width=20pc]{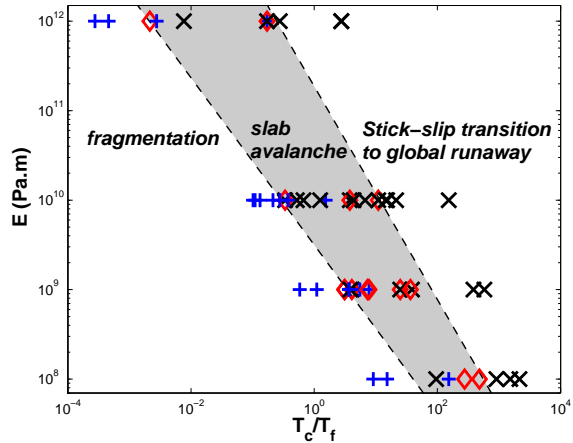}}
\caption{Phase diagram in the plane $[E; T_c/T_f]$ summarizing 
the three main regimes identified. The transition between the 
``stick-slip transition to global runaway'' and the ``fragmentation'' regimes is
continuous. The intermediate regime is characterized by 
the occurrence of a macroscopic slab avalanche (closed to the unstable
behaviour) or global instability (closed to the stable one). The existence
of heterogeneities is found to play an important role in the transitional
behavior.
Fragmentation ($\diamond$), stick-slip ($\times$) and slab avalanche ($+$)
regimes are indicated.
}
\label{phasediag}
\end{figure}

These three regimes are summarized in the phase diagram in the plane
$[E; T_c/T_f]$ shown in Figure \ref{phasediag}.

The elastic coefficient $E$ ($= Y \times l$) controls the  global deformation of the array of block-springs.
Small $E$-values lead to large block displacements as well as 
large deformation of the system. In contrast, for large $E$-values, 
the block displacements are small and the deformations are therefore
small too. A large rigidity favors the emergence of a coherent
behavior of the block motion. $E$ can be understood as controlling
the correlation length of displacements in the system. A larger $E$-value
leads to larger clusters of blocks sliding simultaneously
in local avalanches. The upstream propagation of the stick-slip 
motion is facilitated and develops faster for larger $E$'s.

The ``slab avalanche'' transition regime is found for intermediate values
of $T_c/T_f$, resulting from the competition of instabilities between basal
friction and crack formations. This regime is sensitive to the presence of
heterogeneities. They are modeled by the random resetting of the 
state variable $\theta_i$ of the frictional process after each sliding event.

\section{Conclusion}

With the goal of developing a better understanding of the physical mechanisms leading to accelerated motions
and to increasing seismic activity reported in the case of gravity-driven instabilities, 
we have developed a numerical model based on the discretization
of the natural medium in terms of blocks and springs forming a two-dimensional network
sliding on an inclined plane. Each
block, which can slide, is connected to its four neighbors by springs, that can fail,
depending on the history of displacements and damage. We develop physically
realistic models describing the frictional sliding
of the blocks on the supporting surface and the tensile failure of the springs between blocks proxying
for crack opening. Frictional sliding is modeled with a state-and-velocity
weakening friction law with threshold. Crack formation is modeled with a
time-dependent cumulative damage law with thermal activation including stress corrosion.
In order to reproduce cracking
and dynamical effects, all equations of motion (including inertia) for
each block are solved simultaneously. 

The numerical exploration of the model shows the existence of three regimes as a function 
of the ratio $T_c/T_f$, which are the two characteristic time
scales associated with the two fundamental processes (internal damage/creep
and frictional sliding). For $T_c/T_f \gg 1$, the whole
mass undergoes a series of internal stick and slip events, associated with an initial
slow average downward motion of the whole mass which
progressively accelerates until a global coherent run-away is observed. 
For $T_c/T_f \ll 1$, creep/damage occurs
faster than nucleation of sliding, bonds break and the bottom part of the mass
undergoes a fragmentation process with the creation of a heterogeneous population of 
sliding blocks. For the intermediate regime $T_c/T_f  \sim 1$, a macroscopic crack forms
and propagates along the location of the largest curvature
associated with the slope change. 
In the upper part above the crack, the stable frictional state prevails, while in
the lower part, the unstable frictional sliding state dominates.

Our framework allows for the first time the combination of competing frictional
and damage processes at the origin of mass instabilities. It clears the path to a better understanding of rupture mechanisms in heterogeneous
media. It also casts a gleam of hope for
better forecasting of the ultimate rupture of gravity-driven systems.
Calibrating the relevant physical parameters to different natural materials (soil, rocks, ice or
snow) will also help in the investigation of the slope stability under
external forcing conditions
(such as climate changes). 


\section*{Appendix A: Initiation of sliding for a single block}

This appendix complements 
Section \ref{frictionaa} by providing details of the calculation
of the critical time at which unstable sliding of a given block occurs.
Section \ref{frictionaa} describes the 
sub-critical sliding process of a given block interacting via state-and-velocity
solid friction with its inclined basal surface. When the sub-critical
sliding velocity $d\delta/dt$ diverges (we refer to the time
when this occurs as the ``critical time'' $t_f$ for the frictional sliding instability), 
this signals a change of regime
to the dynamical sliding regime where inertia (the block mass and its acceleration
in the Newton's law) has to be taken into account.

Let us calculate explicitly how the critical time $t_f$ is obtained and define its dependence on the parameters and boundary conditions.
Let us call $T \equiv \Arrowvert \sum \vec F_{\rm bond}\;-\;T_{\rm weight} \vec x \Arrowvert$ 
(or $N \equiv N_{\rm weight}$) the total shear (or normal) force exerted
on a given block, where $\vec F_{\rm bond}$ is the force exerted by a neighboring 
spring bond, and $N_{\rm weight}$ and $T_{\rm weight}$ are the 
normal and tangential forces due to the weight of the block. We then have
\begin{equation}
\label{hjgjhnt'en}
\mu = \frac{T}{N}=\tan \phi ~,
\end{equation}
where $\phi$ is the angle of the basal surface supporting the block.
Therefore, 
\begin{equation}
 \mu_0 + A\;
\ln \frac{\dot{\delta}}{\dot{\delta}_0}+B\; \ln \frac{\theta}{\theta_0} =  \tan \phi ~.
\end{equation}
As explained in Section \ref{frictionaa}, $A -B$ is usually very small for natural material, of the order of $A-B
\approx \pm 0.02$.
For the sake of simplicity, we assume $A=B$. As recalled
in Section \ref{frictionaa}, this choice is not restrictive as it recovers
the two important regimes [Helmstetter et al., 2004]. This leads to
\begin{equation}
\ln \; \frac{\dot{\delta}}{\dot{\delta_0}}\frac{\theta}{\theta_0}\;=\; \frac{\tan\phi-\mu_0}{A}
\end{equation}
whose solution is
\begin{equation}
\dot{\delta}.\theta\;=\;(\dot{\delta}_0\;\theta_0)\;\exp \left(\frac{\tan\phi-\mu_0}{A}\right)~.
\label{kgkgtw}
\end{equation}

From Equation (\ref{qwertyu}), we have
\begin{equation}
\dot{\theta}=1-\frac{\theta \;\dot{\delta}}{D_c}=1- {\;\dot{\delta}_0\;\theta_0 \over D_c}~
\exp \left(\frac{\tan\phi-\mu_0}{A}\right)
\end{equation}
Then
\begin{equation}
\label{thetaev}
\theta\;=\; \theta_0 + \Big( 1-\frac{\;(\dot{\delta}_0\;\theta_0)\;\exp \left(\frac{\tan\phi-\mu_0}{A}\right)}{D_c}\Big) t
\end{equation}
and using Equation (\ref{kgkgtw}), we obtain
\begin{equation}
\dot{\delta} ~=~ { \;(\dot{\delta}_0\;\theta_0)\;\exp \left(\frac{\tan\phi-\mu_0}{A}\right)
\over
\theta_0 + \Big( 1-{\;\dot{\delta}_0\;\theta_0 \over D_c}\;\exp \left(\frac{\tan\phi-\mu_0}{A}\right)\Big) t}~.
\label{ghnt}
\end{equation}
This expression exhibits the usual regimes: a
finite time singularity is obtained for $\;(\dot{\delta}_0\;\theta_0)\;\exp \left(\frac{\tan\phi-\mu_0}{A}\right)>D_c$. 
In this case, Expression (\ref{ghnt}) can be re-written as
\begin{equation}
\label{tfev}
\dot{\delta}\;= {D_c\; \dot{\delta}_0\;\theta_0\;\exp \left(\frac{\tan\phi-\mu_0}{A}\right)
\over D_c - \dot{\delta}_0\;\theta_0\;\exp \left(\frac{\tan\phi-\mu_0}{A}\right)} \cdot {1 \over t_f-t} 
\end{equation}
with
\be
t_f\;=\; {D_c \theta_0
\over D_c - \dot{\delta}_0\;\theta_0\;\exp \left(\frac{\tan\phi-\mu_0}{A}\right)}~.
\ee

We can simplify Expression (\ref{tfev}) by using the condition that, for $\mu=\tan \phi=\mu_0$, we should have $t_f \rightarrow \infty$.
But, for $\mu=\mu_0$, $\exp\left(\frac{\mu_s-\mu_0}{A}\right)=1$ and thus, 
for the condition $t_f \to \infty$ to hold, we need 
\be
{\dot{\delta}_0\;\theta_0 \over  D_c} = 1~.
\label{hntbkwg}
\ee 
The final expression for the critical time $t_f$ signaling the transition from 
a subcritical sliding to the dynamical inertial sliding is, for $\mu>\mu_0$,
\be
t_f\;=\;\frac{\theta_0}{\exp(\frac{\mu_s-\mu_0}{A})-1},
\label{tkhntrk}
\ee
while $t_f \to \infty$ for $\mu \leq \mu_0$. Note that
the dependence on $\dot{\delta_0}$ has disappeared due to the 
relation (\ref{hntbkwg}).

To summarize, a given configuration of blocks and spring tensions determines
the value of  $T \equiv \Arrowvert \sum \vec F_{\rm bond}\;-\;T_{\rm weight} \vec x \Arrowvert$ 
and $N \equiv N_{\rm weight}$ and therefore of $\mu$ via 
(\ref{hjgjhnt'en}). Knowing $\mu$ and given the other material parameters
$\theta_0, \mu_0$ and $A$, we determine the time $t_f$ for the transition to the dynamical
regime for that block via Equation (\ref{tkhntrk}).



\section*{Appendix B: Determination of the rupture time $t_c$ for a single bond}

\subsection*{Case of a constant applied stress up to rupture}

Combining Equations (\ref{eyring}), (\ref{kthnmb}) and (\ref{mgm,tbl;})
in section \ref{hjjbgmfqq} yields
\begin{equation}
\label{geneyring}
\frac{de}{dt} = K \sinh \Bigg( \frac{\beta\; s\;}{ e_{01}^{\mu}}(e+e_{02})^{\mu}\;-\;\beta E e \Bigg)
\end{equation}
for the equation governing the deformation $e$ of a bond subjected to an applied stress $s$.

Nechad et al. [2005] have studied both analytically and numerically the solution of
this equation (\ref{geneyring}) in the case $s>s^*$ (where $s^*$ is defined in the main text)
for which the deformation $e$ blows up to infinity in finite time $t_c$, reflecting the
accelerated tertiary creep regime culminating in the global rupture of the bond.

Our purpose here is to provide a simple approximate formula for the rupture time $t_c$, which
will be used in numerical simulations of the global dynamics of the block array.

An initial method to obtain $t_c$ uses an approximate expression interpolating between the time evolution of $e(t)$ far from and close to $t_c$ found by Nechad et al. [2005]. We propose the expression
\begin{equation}
\label{etest}
e(t)\;=\;\mathcal{A}\;ln(\mathcal{B}+\mathcal{C}t)\Big[-\ln(\frac{t_c-t}{\tau})\Big]^\frac{1}{\mu}\;\;,
\end{equation}
where $\mathcal{A}$,$\mathcal{B}$, $\mathcal{C}$, $t_c$ and $\tau$ are constants
to be determined from matching conditions far from and close to $t_c$.
In particular, Nechad et al. [2005] shows that for \textbf{ $t \ll t_c$}, 
Equation (\ref{geneyring}) simplifies into
\begin{equation}
\label{t<tc}
\frac{de}{dt}\;=\;\frac{a}{b+ct}
\end{equation}
with $a = K e_{02}$, $b = 2e_{02}\;\exp(-\beta s (\frac{e_{02}}{e_{01}})^\mu)$
and $c = K \beta \; (Ee_{02}-\mu s (\frac{e_{02}}{e_{01}})^\mu)$, while 
for \textbf{$t \approx t_c$}, expression (\ref{geneyring}) simplifies into
\begin{equation}
\label{t=tc}
\frac{de}{dt}\;=\;\frac{d}{\mu}\;\Big[-\ln(t_c-t)
\Big]^{(1/\mu)-1}\;\frac{1}{t_c-t}
\end{equation}
with $d\;=\;e_{01}(\beta s)^{-1/\mu}$.

We expand expression (\ref{geneyring}) for $t \ll t_c$and for $t \to t_c$ and 
identify with Equations (\ref{t<tc}) and (\ref{t=tc}) respectively to obtain the system
of five equations for the five unknown $\mathcal{A}$,$\mathcal{B}$,
$\mathcal{C}$, $t_c$ and $\tau$:
\begin{equation}
\left\{\begin{array}{rcl}
e_0 &=&\mathcal{A}\;\ln\mathcal{B}\;\Big(-\ln
\frac{t_c}{\tau}\Big)^\frac{1}{\mu}\;\;,\\
&&\qquad \qquad\qquad\qquad{\rm for}\;\; t \ll t_c\;\;\; e(t=0)=e_0 \\
\frac{a}{b}&=&\mathcal{A}\;\Big(-\ln
\frac{t_c}{\tau}\Big)^{\frac{1}{\mu}-1}\;\Bigg(\mathcal{B}\mathcal{C}\Big(-\ln
\frac{t_c}{\tau}\Big)+\frac{\tau \ln\mathcal{B}}{\mu t_c}\Bigg)\;\;,\\
&&\qquad\qquad\qquad\qquad{\rm for} \;\; t \ll t_c\;\;\; t^0\\
-\frac{ac}{b^2}&=&\mathcal{A}\;\Big(-\ln
\frac{t_c}{\tau}\Big)^{\frac{1}{\mu}-1}\;\Bigg(-\mathcal{C}^2\Big(-\ln
\frac{t_c}{\tau}\Big)+\frac{\tau \mathcal{C}}{\mu t_c
  \mathcal{B}}\Bigg)\;\;,\\
&&\qquad\qquad\qquad\qquad{\rm for} \;\; t \ll t_c\;\;\; t^1\\
\frac{ac^2}{b^2}&=&\mathcal{A}\;\Big(-\ln
\frac{t_c}{\tau}\Big)^{\frac{1}{\mu}-1}\;\Bigg(\frac{\mathcal{C}^2}{\mathcal{B}}\Big(-\ln
\frac{t_c}{\tau}\Big)-\frac{\tau \mathcal{C}^2}{2 \mu t_c
  \mathcal{B}^2}\Bigg)\;\;,\\
&&\qquad\qquad\qquad\qquad{\rm for} \;\; t \ll t_c\;\;\; t^2\\
\frac{d \tau}{\mu t_c}&=&\mathcal{A}\;\Bigg[ \frac{\mathcal{C}}{\mathcal{B}+\mathcal{C}t_c}\Big(-\ln
\frac{t_c}{\tau}\Big)+\frac{\tau\;\ln(\mathcal{B}+\mathcal{C}t_c}{\mu t_c}
\Bigg]\;\;,\\
&&\qquad\qquad\qquad\qquad{\rm for} \;\; t \approx t_c\;\;\; (t_c-t)^0.
\end{array} \right.
\end{equation}

While this system should allow us in principle to determine $t_c$, its nonlinearity
makes it unwieldy. We propose an alternative approach, which consists 
in first simplifying Equation
(\ref{etest}) for $t \ll t_c$ and $t\approx t_c$:
\begin{itemize}
\item For $t \ll t_c$:
\begin{equation}
\label{de<tc}
\frac{de}{dt}\;=\;
\frac{\mathcal{A}\mathcal{C}}{\mathcal{B}+\mathcal{C}t}\;\Bigg[-\ln\Big(\frac{t_c}{\tau}\Big)\Bigg]^{\frac{1}{\mu}}
\end{equation}
\item for $t \approx t_c$:
\begin{equation}
\label{de=tc}
\frac{de}{dt}\;=\;
\mathcal{A}\;\ln(\mathcal{B}+\mathcal{C}t_c)\frac{1}{\mu}\;.\;\frac{1}{t_c-t}\Bigg[-\ln\Big(\frac{t_c-t}{\tau}\Big)\Bigg]^{\frac{1}{\mu}-1}
\end{equation} 
\end{itemize}
In this approach, it is consistent to put $\tau$ equal to $\frac{1}{K}$ where $K$ is the constant defined in equation (\ref{eyring}). Identification with equations (\ref{t<tc}) and (\ref{t=tc}) gives the following
system:
\begin{equation}
\left\{\begin{array}{rcl}
\label{system}
a&=&\mathcal{A}\mathcal{C}\;\Big[-\ln\frac{t_c}{\tau}\Big]^{\frac{1}{\mu}}\;\;,\\
b&=&\mathcal{B}\;\;,\\
c&=&\mathcal{C}\;\;,\\
d&=&\mathcal{A}\;\ln(\mathcal{B}+\mathcal{C}t_c)\;\;.
\end{array} \right.
\end{equation}
Eliminating $\mathcal{A}$ in (\ref{system}) yields an implicit equation that
determines $t_c$:
\begin{equation}
\label{tcev}
\left[-\ln\frac{t_c}{\tau}\right]^{1/\mu}\;\frac{1}{\ln(b+ct_c)}=\frac{a}{cd}
\end{equation}
This implicit equation is found to be sensitive to the value of the initial
parameters and its use is still
rather computer-intensive, which makes it inconvenient for simulations
involving large systems with many blocks.

We have thus turned to a simpler but more approximate method, using the 
asymptotic analytical results obtained in Nechad et al. [2005] that, for $s \gg s^*$
where 

\be
\label{hnnbn}
s^\star=E\;\;\Big(\frac{e_{01}}{\mu}\Big)^\mu\;\;\Big(\frac{\mu-1}{e_{02}}\Big)^{\mu-1}
\ee
is the minimum stress such that the material will fail eventually, 
$t_c$ is given by 
\be
t_c = C  \;\exp(-\gamma \;s)~,
\label{tkhtnh2t}
\ee
where
\be
\gamma=\frac{\beta e_{02}^\mu}{e_{01}^\mu}~.
\label{hy3ghte}
\ee
As $K$ has the dimension of the inverse of time in Equation
(\ref{geneyring}), we take the constant $C$ to be proportional to $1/K$: $C \sim
\frac{1}{K}$. Our formula for determining $t_c$ in our simulations is therefore
summarized by the following equations:
\begin{equation}
\label{tcs}
t_c =\left\{ 
\begin{array}{ll}
{1 \over K} \;\exp(-\gamma \;s) & \textrm{if } s > s^\star\\
\to \infty & \textrm{if }  s \leq s^\star
\end{array} \right.
\end{equation}

\section{Appendix C: Energy analysis}
In order to analyze our results in term of icequake activity 
and compare numerical with experimental results, we estimate the energy
evolution in the lattice.
Here, four types of energy can be distinguished and evaluated: 
gravitational potential energy, kinetic energy, energy
stored in the elastic bonds (analogous to an internal energy) and radiated energy. 

\paragraph{Gravitational potential energy}

The gravitational potential energy is evaluated using the classical expression:
\begin{equation}
E_p(t)\;=\;\sum_{all \;\;blocks} m g \;\;z_{block}
\end{equation}

\paragraph{Energy stored in bonds} For each bond linking block $i$
and block $j$, the stored elastic energy is defined by:
\begin{equation}
E_b(t) = \frac{1}{2}\sum_{all\;\;bonds} E \Big(b_{ij} \Arrowvert \vec x_j-\vec x_i
\Arrowvert-l \Big)^2
\end{equation}
where $\vec x_j$ and $\vec x_i$ are the position of two neighboring blocks and
$l$ the initial length of each bond.

\paragraph{Kinetic energy}
The total kinetic energy of the system is evaluated according to the formula
\begin{equation}
E_k(t) = \sum_{all\;\;blocks} \frac{1}{2} m \Big(\frac{\Arrowvert \vec x(t+\delta t)-\vec x(t)\Arrowvert}{\delta
t}\Big)^2
\end{equation}
with $\vec x(t)$ the position of the considered block at time $t$.

\paragraph{Radiated energy}
The radiated energy $E_r(t)$ between two time steps is
evaluated using:
\begin{equation}
E_r(t)\;=\;E_t-E_p(t)-E_b(t)-E_k(t)
\end{equation}
with $E_t=E_p(0)+E_b(0)+E_k(0)=E_p(0)$. In other words,
it is the additional energy not found in the potential, elastic 
or kinetic energies.

\pagebreak

\end{document}